\documentclass[%
 reprint,
superscriptaddress,
 showkeys,
 amsmath,amssymb,
 aps,
]{revtex4-2}

\usepackage{graphicx}
\usepackage{dcolumn}
\usepackage{bm}
\usepackage[textsize=tiny]{todonotes}

\renewcommand{\eqref}[1]{Eq.~(\ref{#1})}
\newcommand{\figref}[1]{Fig.~\ref{#1}}

\newcommand{\kn}{k_{\text{n}}}
\newcommand{\taug}{\tau_{\text{g}}}
\newcommand{\Vdrop}{V_{\text{drop}}}

\newcommand{\SIfigSecondary}{S1}
\newcommand{\SIfigSinglePoly}{S2}

\newcommand{\SIfigTaugExample}{S12}
\newcommand{\SIfigTaug}{S13}
\newcommand{\SIfigStepwise}{S14}
\newcommand{\SIfigStepwiseDetails}{S15}
\newcommand{\SIfigSeeded}{S16}
\newcommand{\SIfigBulk}{S17}
\newcommand{\SIfigCellphone}{S18}

\begin{document}

\title{Macroscopic DNA-programmed photonic crystals via seeded growth}
\author{Alexander Hensley}
\affiliation{%
 Martin A. Fisher School of Physics, Brandeis University, Waltham, MA 02453 USA\\
}%
\author{Thomas E. Videb\ae k}
\affiliation{%
 Martin A. Fisher School of Physics, Brandeis University, Waltham, MA 02453 USA\\
}%
\author{Hunter Seyforth}
\affiliation{%
 Martin A. Fisher School of Physics, Brandeis University, Waltham, MA 02453 USA\\
}%
\author{William M. Jacobs}%
 \email{wjacobs@princeton.edu}
\affiliation{%
 Department of Chemistry, Princeton University, Princeton, NJ 08544 USA\\
}%
\author{W. Benjamin Rogers}%
 \email{wrogers@brandeis.edu}
\affiliation{%
 Martin A. Fisher School of Physics, Brandeis University, Waltham, MA 02453 USA\\
}%

\date{\today}

\begin{abstract}
  Photonic crystals---a class of materials whose optical properties derive from their structure in addition to their composition---can be created by self-assembling particles whose sizes are comparable to the wavelengths of visible light.
  Proof-of-principle studies have shown that DNA can be used to guide the self-assembly of micrometer-sized colloidal particles into fully programmable crystal structures with photonic properties in the visible spectrum.
  However, the extremely temperature-sensitive kinetics of micrometer-sized DNA-functionalized particles has frustrated attempts to grow large, monodisperse crystals that are required for photonic metamaterial applications.
  Here we describe a robust two-step protocol for self-assembling single-domain crystals that contain millions of optical-scale DNA-functionalized particles: Monodisperse crystals are initially assembled in monodisperse droplets made by microfluidics, after which they are grown to macroscopic dimensions via seeded diffusion-limited growth.
  We demonstrate the generality of our approach by assembling different macroscopic single-domain photonic crystals with metamaterial properties, like structural coloration, that depend on the underlying crystal structure. By circumventing the fundamental kinetic traps intrinsic to crystallization of optical-scale DNA-coated colloids,  we eliminate a key barrier to engineering photonic devices from DNA-programmed materials.
\end{abstract}

\keywords{Photonic crystals | Crystallization | Self-assembly | DNA-programmed colloids}

\maketitle

\section{Introduction}
DNA-programmed self-assembly leverages the chemical specificity of DNA hybridization to stabilize user-prescribed crystal structures~\cite{Rogers2016,laramy2019crystal}.
Pioneering studies have demonstrated that DNA hybridization can guide the self-assembly of a wide variety of nanoparticle crystal lattices, which can grow to micrometer dimensions and contain millions of particles~\cite{mirkin1996dna, Mirkin2008Nature, Oleg2008Nature, macfarlane2011nanoparticle, zhang2013general, auyeung2014dna, Mirkin2014Nature, liu2016diamond}.
Attention has now turned toward the goal of assembling \textit{photonic crystals} from \textit{optical-scale} particles~\cite{yablonovitch1994photonic,joannopoulos1997photonic,he2020colloidal,sun2018design} using DNA-programmed interactions.
To this end, progress over the past decade has established that DNA can indeed program the self-assembly of bespoke crystalline structures from micrometer-sized colloidal particles~\cite{Crocker2012NComm,rogers2015programming,Wang2015,Wang2017NComm,ducrot2017colloidal,fang2020two,Hensleye2114050118}.
However, growing single-domain crystals comprising millions of DNA-functionalized, micrometer-sized colloidal particles remains an unresolved barrier to the development of practical technologies based on DNA-programmed assembly.
Prior efforts have yielded either single-domain crystals no more than a few dozen micrometers in size~\cite{Crocker2012NComm,rogers2015programming,Wang2015,Wang2017NComm} or larger polycrystalline materials with heterogeneous domain sizes~\cite{Wang2015,ducrot2017colloidal,oh2020reconfigurable,he2020colloidal}.
These features---small crystal domains, polycrystallinity, and size polydispersity---have therefore precluded the use of DNA-coated colloidal crystal in photonic metamaterial applications. 

\begin{figure*}
    \centering
    \includegraphics[width=\linewidth]{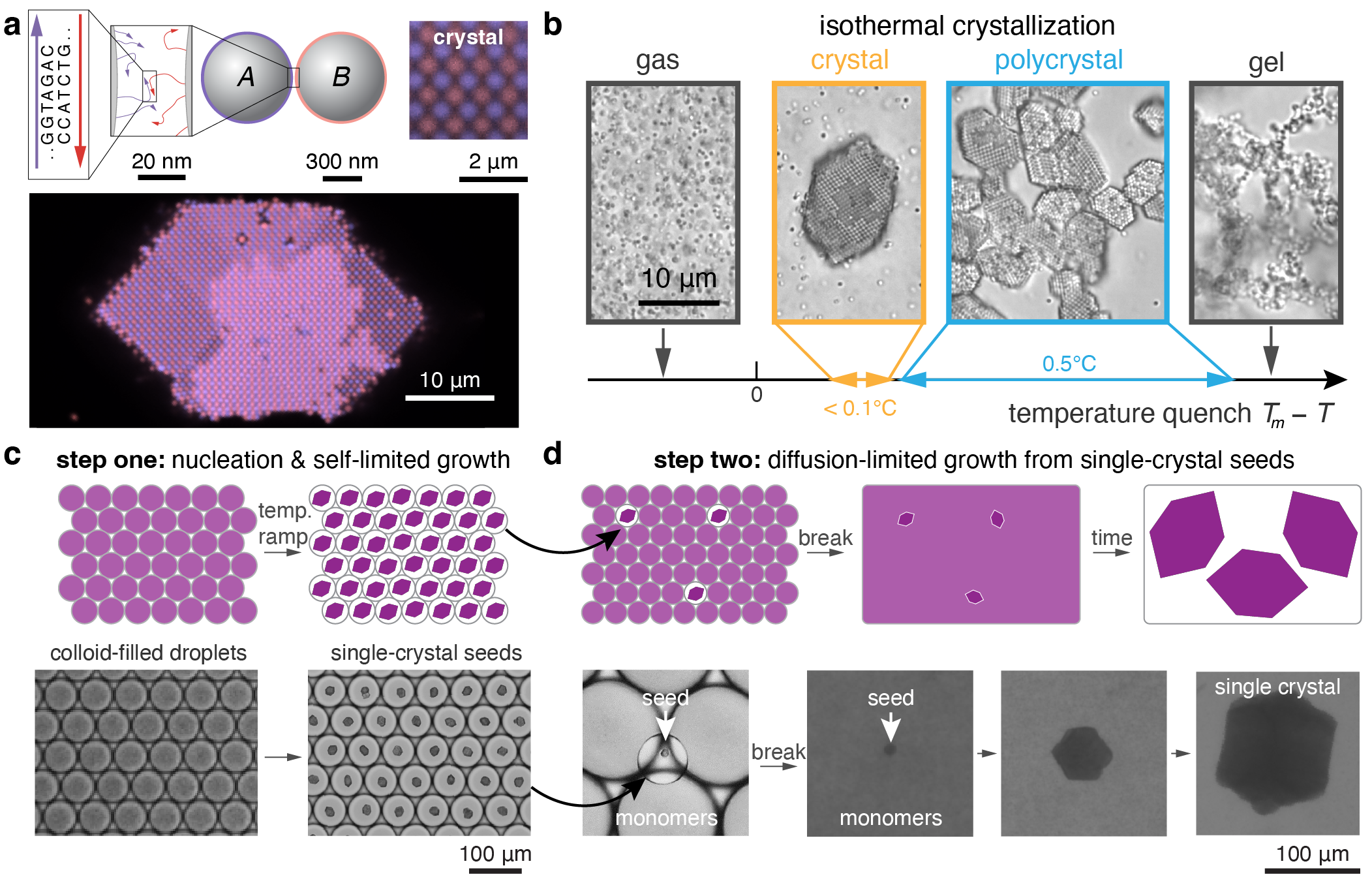}
    \caption{ \textbf{Large single-domain crystals can be self-assembled from DNA-coated colloids via a two-step process.} (\textbf{a}) DNA molecules at the nanometer-scale can link together micrometer-scale colloidal particles to program the assembly of colloidal crystals. The fluorescence micrograph shows a DNA-programmed binary colloidal crystal formed from 600-nm-diameter particles. The inset micrograph shows the crystal lattice. (\textbf{b}) DNA-programmed crystallization is strongly temperature dependent. Below the melting temperature, $T_{\text{m}}$, single crystals can form in a very narrow temperature window, indicated in orange. Just below this temperature window, kinetically arrested polycrystalline or gel-like assemblies form. Optical micrographs show examples of these different states for the same particles as in \textbf{a}. (\textbf{c}) The first step of our protocol involves nucleating size-monodisperse single crystals in monodisperse water droplets made via microfluidics, shown in a cartoon schematic (top) and in brightfield micrographs (bottom). Monodisperse droplets filled with DNA-coated colloids are slowly cooled to produce same-size single crystals. (\textbf{d}) The second step involves recovering the single crystals by breaking the emulsion and then using them to seed crystal growth in a metastable colloidal suspension, shown in a schematic (top) and in brightfield micrographs (bottom). A small number of crystal-containing droplets from \textbf{a} are combined with droplets containing DNA-coated particles. The emulsion is ruptured and the system is cooled to a temperature at which crystals grow but nucleation of additional crystals is suppressed.}
    \label{fig:intro}
\end{figure*}

Assembling macroscopic materials from DNA-functionalized, micrometer-sized colloids is challenging due to the vastly different length scales between the DNA molecules and the colloidal particles (\figref{fig:intro}a).
This combination leads to crystallization kinetics that are extremely sensitive to temperature~\cite{Rogers2016} and prone to kinetic trapping~\cite{dreyfus2009simple,rogers2011direct,cui2022comprehensive}.
The resulting challenges are both practical and fundamental in nature.
For example, recent work has shown that crystal nucleation rates can vary by orders of magnitude over a temperature range of only $0.25^\circ${C}~\cite{Hensleye2114050118}.
Extremely precise temperature control would therefore be required to self-assemble single-domain crystals from a bulk solution (\figref{fig:intro}b).
At the same time, annealing polycrystalline materials is difficult due to the combination of the short-range attraction and the friction arising from the DNA-mediated colloidal interactions, which slows the rolling and sliding of colloidal particles at crystalline interfaces~\cite{Wang2015,Miranda2018SoftMatter,jana2019translational,Hensleye2114050118}.
Thus, the impracticality of annealing crystals composed of DNA-functionalized, micrometer-sized particles results from intrinsic features of these materials that cannot easily be designed around.

In this Article, we describe a two-step approach to overcome these limitations and self-assemble macroscopic photonic crystals from DNA-coated colloidal particles.
We first show that small, monodisperse single crystals can be reliably assembled within nanoliter-scale microfluidic droplets by subjecting the droplets to a simple cooling protocol (Fig.~\ref{fig:intro}c).
Then, to go beyond a fundamental size limitation imposed by droplet-confined assembly, we use these crystals to seed continued growth in a bulk solution, while simultaneously suppressing further crystal nucleation (Fig.~\ref{fig:intro}d).
We develop a theoretical framework that models both processes quantitatively, enabling us to make monodisperse suspensions of single crystals with prescribed dimensions and predictable yields.
Finally, by varying the size of the colloidal particles and the duration of the secondary growth phase, we show that our approach can be easily generalized to create a variety of monodisperse macroscopic crystals with different crystal structures and therefore tunable photonic properties, including crystals that can be seen by the naked eye.
Most importantly, we emphasize that this approach for synthesizing macroscopic crystals is \textit{robust}, meaning that the procedure is insensitive to small changes in processing conditions, can be repeated over and over again, and can be applied to different particle sizes and compositions. Therefore, our platform could enable significant future advances in DNA-programmed assembly of both nanometer- and micrometer-scale colloids. 

\section{Results}
\subsection{Assembly of monodisperse crystalline seeds}
We first seek to understand the physics governing the self-assembly of colloids confined to small water droplets subject to a time-dependent cooling protocol (Step One; \figref{fig:intro}c). Whereas we previously demonstrated that small single-domain colloidal crystals could be assembled in monodisperse water droplets ~\cite{Hensleye2114050118}, here we aim to 
understand how to optimize the yield of single-domain crystals assembled via this approach.
We thus perform systematic experiments for a particular prototypical crystal structure, isostructural to copper gold (CuAu)~\cite{Hensleye2114050118}, to validate a parameter-free analytical theory, which provides insights into the fundamental physical limitations of this method, as we discuss below.

\begin{figure}
    \centering
    \includegraphics{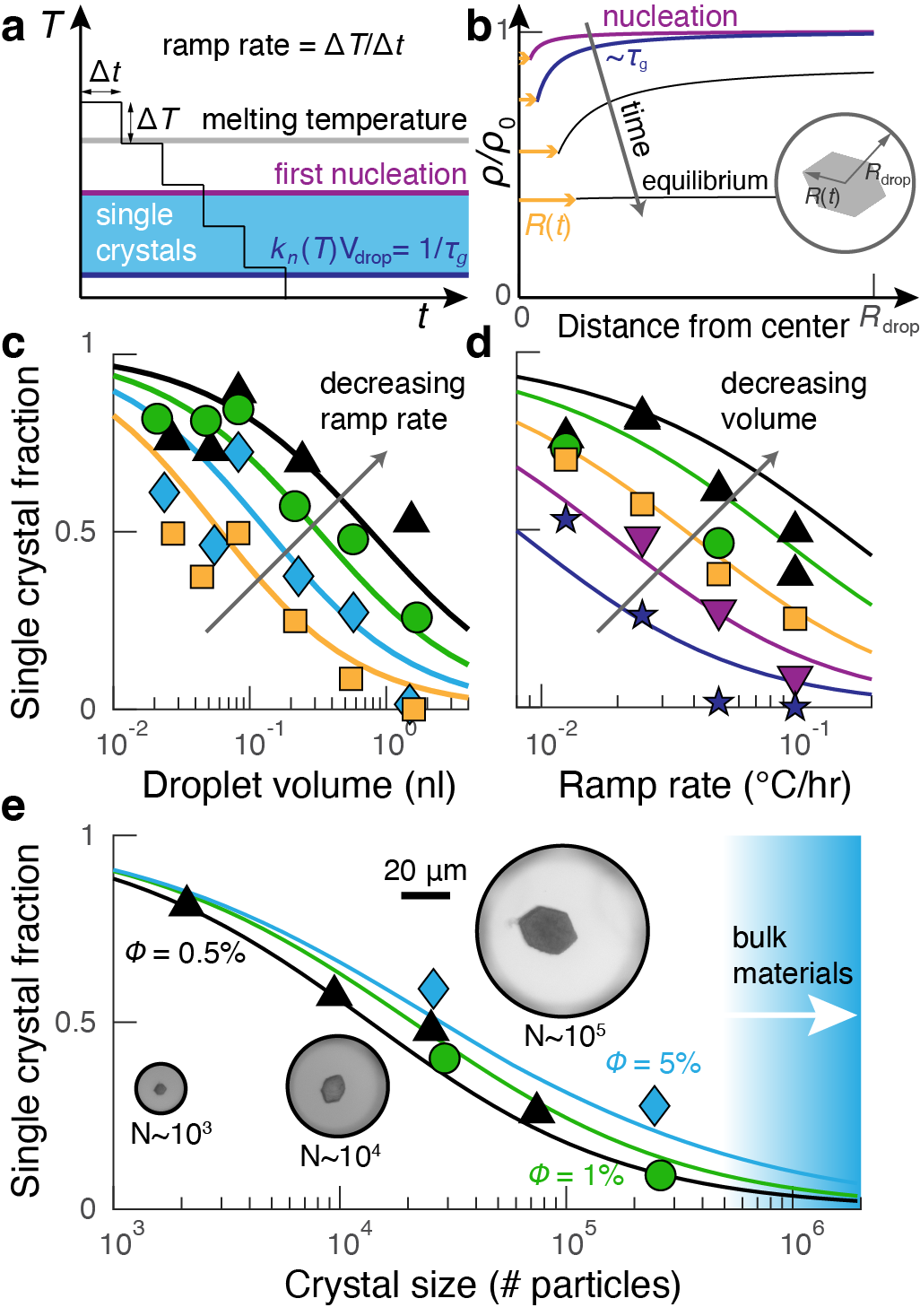}
    \caption{\textbf{Linear cooling produces single crystals within nanoliter droplets.} (\textbf{a}) The temperature ramp is a series of discrete $0.1^\circ$C steps with a duration of $\Delta t$, which starts above the melting temperature and runs until all of the particles are incorporated into the crystals. (\textbf{b}) The relative colloid concentration as a function of the distance from the center of the crystal at different times during crystal growth. Curves show the predicted concentration profiles and orange arrows indicate the crystal radius, $R(t)$; $\rho$ is the colloid concentration and $\rho_0$ is the uniform concentration at time $t=0$. (\textbf{c}) The fraction of droplets that form single crystals as a function of droplet volume. Points represent experiments and lines show model predictions for different ramp rates:  $0.0125^\circ$C/hr (triangles),  $0.025^\circ$C/hr (circles),  $0.05^\circ$C/hr (diamonds), and $0.1^\circ$C/hr (squares). (\textbf{c}) The data in \textbf{d} replotted versus the ramp rate. Different shapes represent different droplet volumes: $0.04~\mu$l (triangles), $0.08~\mu$l (circles), $0.2~\mu$l (squares), $0.5~\mu$l (diamonds), and $1.4~\mu$l (stars). (\textbf{e}) The fraction of single crystals versus the number of particles within each droplet using a ramp rate of $0.025^\circ$C/hr. Different symbols correspond to different particle volume fractions. The single-crystal fraction decreases monotonically with decreasing particle concentration and increasing crystal size. In all cases, the data and model predictions approach zero before crystals reach sizes of $10^6$ particles, which we define as the threshold for macroscopic materials. The particles are 600 nm in diameter in \textbf{c}--\textbf{e}.}
    \label{fig:ramp}
\end{figure}

Using microfluidics, we create a water-in-oil emulsion of nanoliter-scale droplets that are filled with a binary suspension of 600-nm-diameter DNA-coated polystyrene particles, and subject the droplets to a linear cooling protocol.
Throughout the protocol, the temperature is deceased in a staircase fashion with a specified time delay $\Delta t$ and a temperature change $\Delta T=-0.1^\circ \textrm{C}$ (\figref{fig:ramp}a).
Because the DNA-programmed attractions become stronger with decreasing temperature, one or more crystals eventually nucleate and grow within each droplet (\figref{fig:ramp}b), leading to an ensemble of small crystals throughout the emulsion.
The temperature ramp then continues until all the colloidal particles in each droplet are incorporated into the crystal phase.
At the conclusion of the temperature ramp, we measure the fraction of droplets within the ensemble that contain precisely one single-domain crystal (see Fig.~\SIfigSinglePoly~for an example).
Importantly, because the final crystal phase incorporates all the colloidal particles, single-domain crystals assembled via this technique are monodisperse in volume to within the typical 5\% variation in the initial particle loading~\cite{Hensleye2114050118}.
See Supplementary Information (SI) for experimental details.

We find that the yield of single crystals decreases with increasing droplet volume and increasing ramp rate.
Both of these trends can be rationalized by considering the pathway by which single crystals form: The growth of the first crystal that nucleates must reduce the supersaturation throughout the droplet in order to suppress the nucleation of additional crystals~(\figref{fig:ramp}b).
We therefore consider the factors that influence both the temperature at which the first crystal is most likely to nucleate and the probability of subsequent nucleation events.
We propose that varying the droplet size (\figref{fig:ramp}c) primarily affects the single-crystal fraction by altering the probability of secondary nucleation events.
Because the droplet size has a relatively small influence on the initial nucleation temperature, the likelihood of secondary nucleation is determined by the time required to reduce the supersaturation by particle diffusion. 
Therefore, at a fixed ramp rate, the single-crystal fraction is smaller in larger droplets because the diffusive time scale is proportional to the square of the droplet diameter.
By contrast, we propose that changing the ramp rate (\figref{fig:ramp}d) primarily affects the initial nucleation temperature at which the first crystal is formed within the droplet.
Specifically, at a fixed droplet volume, faster ramp rates strongly bias the initial nucleation event towards lower temperatures as a direct result of the reduced duration of the higher-temperature steps.
Because particle diffusion is comparatively insensitive to temperature, the dominant effect of a faster ramp rate is likely to be the increased instantaneous nucleation rate at the lower first-nucleation temperature, which in turn increases the probability of additional nucleation events.

To describe these factors quantitatively, we introduce the growth time, $\taug$, to describe the typical time required for crystal growth to suppress further nucleation elsewhere in the droplet.
Intuitively, the nucleation rate should be slower than $1/\taug$ in order for a single crystal to assemble (\figref{fig:ramp}a).
For example, in an isothermal protocol, the probability that a second crystal does \textit{not} nucleate is given by $\exp[-\Vdrop\kn(\rho_0,T)\taug]$, where $\Vdrop$ is the droplet volume and $\kn$ is the nucleation rate density at the initial colloid concentration $\rho_0$ and temperature $T$.
Even though the situation is more complicated in the case of a temperature ramp, $\taug$ is conceptually and computationally useful because it is only weakly dependent on temperature (see SI for a formal definition and further details).

Theoretical predictions based on a modified classical theory of nucleation and growth enable us to predict $\taug$ and thus to describe our experimental results quantitatively with no adjustable parameters.
Using previously determined concentration and temperature dependencies of the nucleation rate and crystal growth velocity~\cite{Hensleye2114050118}, we first predict the probability of the initial nucleation event and the ensuing crystal growth dynamics for a prescribed droplet size and temperature protocol.
Importantly, this model captures the evolution of the concentration field around the first crystal that nucleates, allowing us to predict the decrease in the supersaturation throughout the entire droplet volume as this initial crystal grows, which is necessary to compute $\taug$ quantitatively (\figref{fig:ramp}b and Fig.~\SIfigTaugExample).
We then use these calculations to predict the probability of growing a single crystal in a droplet during a time-dependent temperature protocol using a generalization of the isothermal expression given above (Fig.~\SIfigTaug; see SI for details).

Across nearly the full range of parameter space, we find that our model captures our measured single-crystal fractions quantitatively, within the uncertainty of our experimental measurements (\figref{fig:ramp}c-d).
We note that systematic deviations between the predictions and measured single-crystal fraction are observed at the slowest ramp rates, for which the measurements exhibit nonmonotonic dependencies on the ramp rates and droplet volumes. However, we attribute these effects to substantial evaporation of the solvent from the droplets, which can reduce the smallest droplet volumes by as much as 30\% during annealing.
Therefore, the overall accuracy of our model allows us to predict the conditions required to achieve a target yield and to optimize the temperature-ramp protocol subject to a maximum duration of the experiment.

Unexpectedly, our theoretical model predicts that utilizing discrete temperature steps, as opposed to a continuous ramp, is in fact beneficial for maximizing the single-crystal yield at a fixed ramp rate, $|\Delta T / \Delta t|$ (Fig.~\SIfigStepwise; see SI for details).
As long as $\Delta t$ is longer than $\taug$, each discrete step can be considered as an isothermal protocol, which optimally suppresses further nucleation by holding the nucleation rate density, $\kn$, constant for the entirety of $\taug$.
By contrast, a continuous ramp implies that $\kn$ increases continuously following the first nucleation event, increasing the probability of secondary nucleation events.
Nonetheless, $\Delta t$ cannot be made too large, as the correspondingly large temperature steps required to maintain the fixed ramp rate would tend to bias the first nucleation event to lower temperatures, and thus nucleation rates that are faster than $1/\taug$.
Balancing these competing factors, our model predicts that the single-crystal probability is maximized for temperature steps on the order of $\Delta T = 0.1^\circ${C}.
As this is the step size used in our experiments, our modeling suggests that further refinement of the precise functional form of our temperature protocol would yield minimal improvement (Fig.~\SIfigStepwiseDetails).
In practice, we therefore only need to tune the step duration, $\Delta t$, to achieve a target single-crystal yield using a prescribed droplet volume and particle concentration.

Despite its many advantages, the temperature-ramp protocol is not a magic bullet for assembling \emph{macroscopic} photonic crystals from DNA-coated colloids, since both theory and experiment point to fundamental physical limitations on the size of single crystals that can be assembled in droplets at high yield.
Since larger droplet volumes reduce the yield, growing larger crystals at a fixed ramp rate can only be achieved by increasing the particle concentration.
Yet, when comparing different cooling protocols that would yield a given final crystal size, we find that increasing the initial colloidal concentration to as much as $5\%$ by volume has only a modest effect on the single-crystal yield (Fig.~\ref{fig:ramp}e).
On the other hand, we encounter a practical limitation when reducing the ramp rate below $0.025^\circ$C/hr, since extremely long annealing protocols are accompanied by substantial evaporation of the solvent from the droplets as noted above.
Thus, taken together, our model and experiments demonstrate that single-domain crystals containing more than one million DNA-functionalized, micrometer-sized particles cannot be assembled in droplets within a practical annealing duration with any appreciable yield, regardless of the precise form of the temperature protocol.

\begin{figure}
    \centering
    \includegraphics{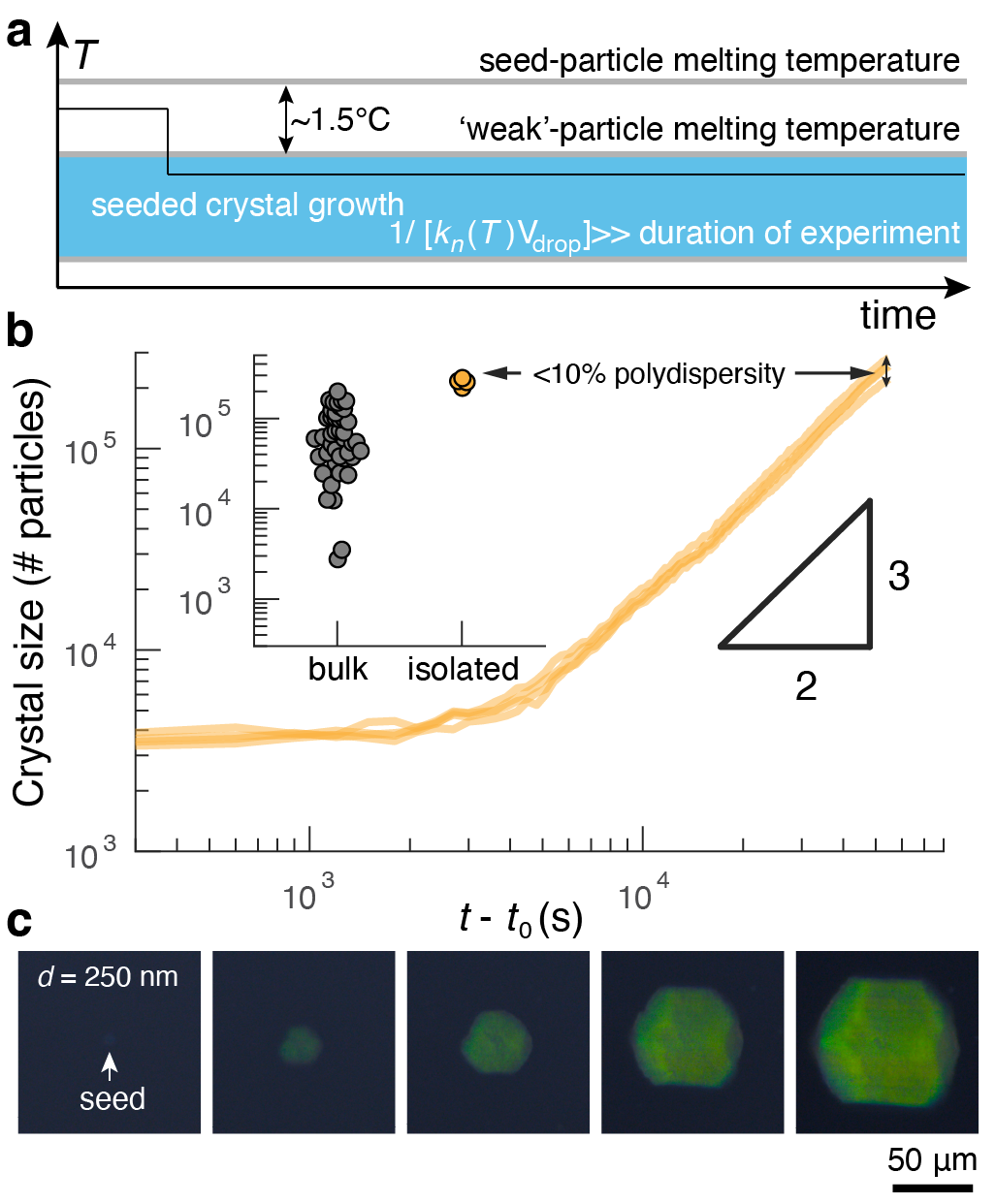}
    \caption{\textbf{Seeded crystal growth is diffusion limited and yields monodisperse large crystals.} (\textbf{a}) The temperature protocol for the seeded growth step. The supercooling is kept roughly constant during the experiment such that the seeds can grow but homogeneous nucleation is suppressed.   (\textbf{b}) A plot of the crystal size of four seeds in the same sample over time. The particles are 600 nm in diameter. We define $t_0$ to be the time at which the crystal has grown by ten percent in size. In all cases, the growth is diffusion limited with a characteristic $3/2$ power scaling. In this specific experiment, the crystals are spaced apart by roughly four times their final diameter. The inset shows the crystal sizes at the end of the experiment for seeded growth (orange) and bulk nucleation and growth (gray). (\textbf{c}) Micrographs of seeded growth of 250-nm-diameter particles imaged in transmission through crossed polarizers shows the appearance of coloration upon growth.}
    \label{fig:growth}
\end{figure}

\subsection{Diffusion-limited seeded growth}
To overcome this fundamental limitation, we introduce a second processing step in which the small, monodisperse single crystals assembled in droplets can be grown by orders of magnitude in size (Step Two in \figref{fig:intro}d).
This second step exploits the fact that particles are able to attach to a crystal surface at a higher temperature, and thus lower supersaturation, than that at which nucleation occurs.
To this end, we break the emulsion and transfer the droplet-assembled crystals to a fresh bath of `weak' particles whose DNA grafting density is reduced by half during particle synthesis. The resulting lower melting temperature of the weak particles allows them to exist in the gas phase at temperatures below the melting temperature of the seed crystals themselves.
We then anneal the system at a temperature above the homogeneous nucleation temperature of the weak particles but below the melting temperature of the crystalline seed (Fig.~\ref{fig:growth}a and Fig.~\SIfigSecondary).
Because the seed crystal size itself is not essential, we choose seeds from a slow ramp protocol that uses a ramp rate of $0.1^\circ$C/4~hr and a droplet volume of roughly 0.05~nl, resulting in a single-crystal fraction of nearly 100\%.

We track the size of single-crystalline seeds during secondary growth by measuring their area as a function of time and then relating the two-dimensional area to the number of particles per crystal, $N(t)$, using an empirical calibration curve (Fig.~\ref{fig:growth}b).
Since we expect that seeds will be effectively isolated from one another if their diffusion fields do not strongly overlap, we dilute the seeds until they are separated by at least a few multiples of the target crystal diameter at the end of the secondary growth experiment.
Consistent with this expectation, growth curves of isolated seeds confirm that seeded crystals can be grown by orders of magnitude in size and that their size follows the scaling relation predicted by diffusion-limited growth, with $N(t) \propto t^{3/2}$ at long times (Fig.~\ref{fig:growth}b and Fig.~\SIfigSeeded).
Note that at early times, the crystal size does not follow the same power-law scaling because the crystals begin with a finite initial size of roughly 3000 particles per crystal in this experiment (see SI for details on the experiment and calibration).

Importantly, because diffusion-limited growth is deterministic, the size distribution of the final crystals after the secondary growth step retains the monodispersity of the original crystal seeds.
\figref{fig:growth}b shows that the standard deviation of the final crystal volumes is less than 10\% of the mean.
By contrast, the final sizes of colloidal crystals nucleated in a bulk solution vary by roughly two orders of magnitude, reflecting both the distribution of nucleation times as well as competition for free particles from neighboring growing crystals (Fig.~\ref{fig:growth}b, inset, and Fig.~\SIfigBulk).
Thus the deterministic growth behavior in our approach not only allows us to predict the duration of the secondary growth phase required to grow a single-domain crystal of a prescribed size, but it also allows us to make many same-sized macroscopic colloidal crystals in a single experiment.
In principle, there is no upper limit to the size of crystals that can be prepared using our approach; growing even larger crystals simply requires reducing the seed density and increasing the growth time.

To demonstrate the direct connection between the bulk optical properties and the crystal size, we perform the same seeded-growth experiment using particles that are 250 nm in diameter. Figure~\ref{fig:growth}c shows snapshots from a typical growth trajectory. Whereas the seed is almost imperceptible, we see that the crystal exhibits prominent coloration upon growth and that its color saturation increases until it is roughly 50 micrometers in linear dimension. The final snapshot indicates that the final crystal contains millions of particles and is therefore well beyond the maximum size that can be assembled directly in a microfluidic droplet.

\subsection{Macroscopic photonic crystals}

\begin{figure*}[]
    \centering
    \includegraphics[width=\linewidth]{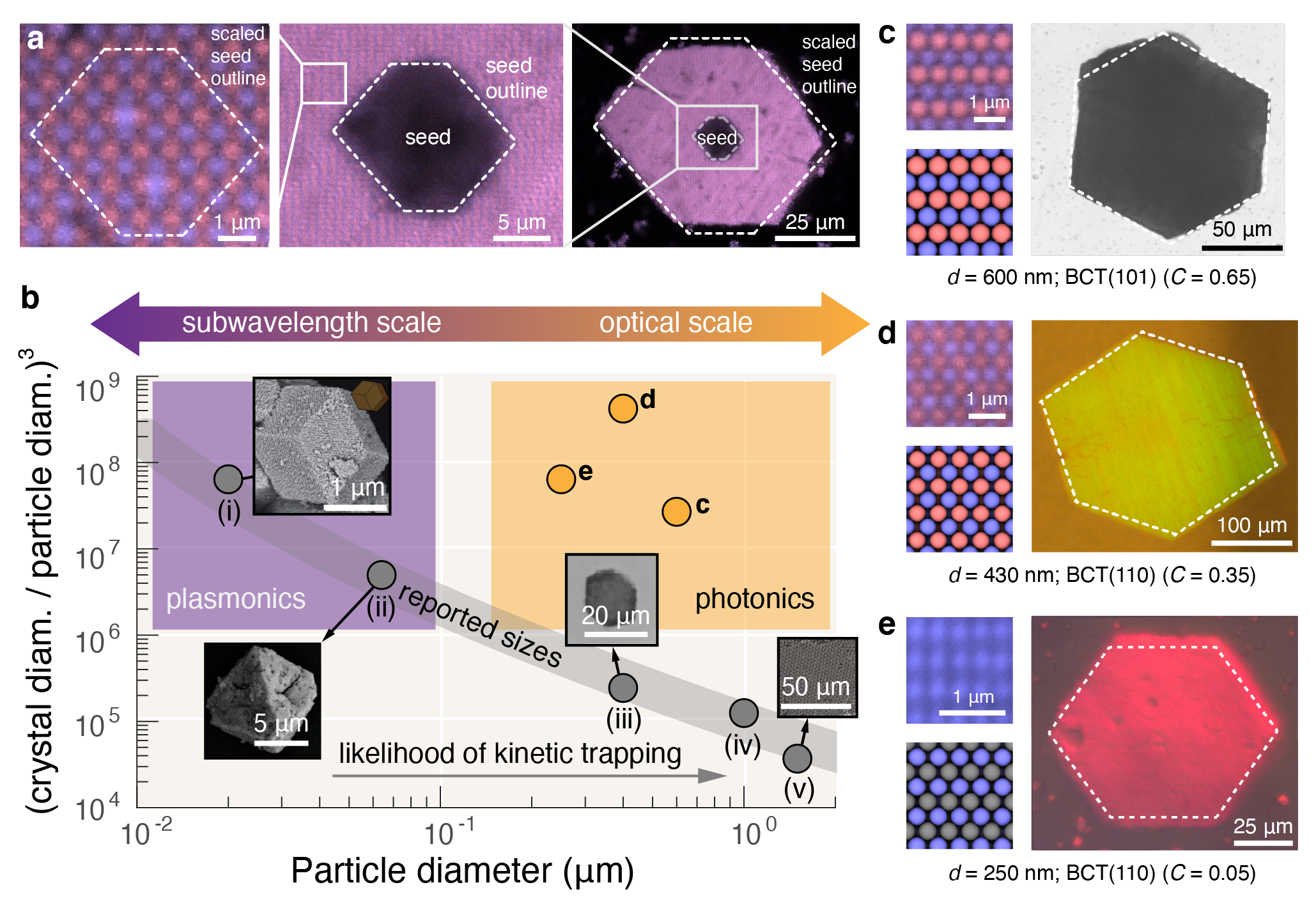}
    \caption{ \textbf{Single crystals from optical-scale particles can grow to macroscopic dimensions and exhibit photonic properties dependent on the crystal structure.} (\textbf{a}) Confocal fluorescence images of a seeded crystal of 600-nm-diameter particles. The crystal has a well-defined habit (right) that is consistent with the seed habit (middle) and the underlying lattice structure of the crystal (left). The seed particles are not dyed. (\textbf{b}) An overview of the reported sizes of the largest crystals of DNA-coated particles from the literature as a function of the constituent particle size, spanning subwavelength- to wavelength-scale particles. Generally, the number of particles per single-domain crystal decreases as the particle size increases due to kinetic trapping (gray points). Our two-step protocol breaks this trend, allowing well-faceted crystals to be grown multiple orders of magnitude larger than before. Orange points show the largest single-crystal sizes that we synthesized for three particle diameters, shown in \textbf{c}, \textbf{d}, and \textbf{e}. (\textbf{c}) A brightfield micrograph of a single-crystal of 600-nm-diameter particles, which has a crystal habit (dashed outline) consistent with the (101) view of a body-centered tetragonal (BCT) crystal structure (BCT parameter $C=0.65$) as shown in the insets: experiment (top) and model (bottom). (\textbf{d}) A micrograph imaged in reflection through crossed polarizers of a single-crystal of 430-nm-diameter particles, which has a crystal habit (dashed outline) consistent with the (110) view of BCT ($C=0.35$), as shown in the insets: experiment (top) and model (bottom). (\textbf{e}) A micrograph imaged in reflection through crossed polarizers of a single-crystal of 250-nm-diameter particles, which has a crystal habit (dashed outline) consistent with the (110) view of BCT ($C=0.05$), as shown in the insets: experiment (top) and model (bottom); we only show one particle species because the other species' dye emits in the red so the particles are below the diffraction limit and cannot be resolved. The crystal structure in \textbf{c} is closest to FCC-CuAu; the structure in \textbf{d} is intermediate between FCC-CuAu and CsCl; and the structure in \textbf{e} is isostructural to CsCl. All crystals were grown for roughly two days. Literature points and corresponding micrographs are from the following references: (i)~\cite{auyeung2014dna}, (ii)~\cite{o2016programming}, (iii)~\cite{Hensleye2114050118}, (iv)~\cite{oh2020high}, and (v)~\cite{he2020colloidal}.}
    \label{fig:crystals}
\end{figure*}

Finally, we turn our attention toward using our two-step platform to assemble macroscopic single-crystalline materials from DNA-coated colloids. To this end, we fist confirm that the seeded crystals are indeed single crystalline by imaging them at high magnification with single-particle resolution.
An example of a crystal assembled from 600-nm-diameter particles is shown in \figref{fig:crystals}a. To demonstrate that the crystal was indeed grown from a seed, we use undyed seed particles and fluorescently labeled `weak' particles. The seed is clearly visible in the interior of the crystal, as is the seed outline or the seed crystal habit (\figref{fig:crystals}a). Zooming in on the crystal lattice, we see the crystalline order with single-particle resolution. Furthermore, we see that a scaled version of the seed outline follows the crystallographic directions of the lattice, indicating that growth preserves the crystal structure of the seed. Similarly, we see that a scaled version of the seed outline also approximates the habit of the large-scale seeded crystal itself. 
Therefore, we conclude that the assembled structure---which contains roughly one million particles---is a single crystal. 

To further highlight the importance of our two-step approach, we plot the estimated crystal sizes from the literature versus the particle diameter to illustrate the challenge that the field of DNA-programmed assembly has faced in making bulk self-assembled materials from optical-scale particles: Whereas the largest single crystals made from 20-nm-diameter DNA-coated colloids contain roughly 100 million particles~\cite{auyeung2014dna,o2016programming}, the kinetic limitations associated with using DNA to form crystals from micrometer-scale particles have restricted such crystals to 1000-fold fewer particles per crystallite~\cite{Hensleye2114050118,oh2020high,he2020colloidal}.
A direct consequence of this hurdle is that, prior to our work, the region of parameter space corresponding to bulk photonic crystals was completely vacant (\figref{fig:crystals}b, orange box). 

Our approach enables us to populate this space and create macroscopic photonic crystals using DNA-programmed self-assembly.
We demonstrate the potential of our experimental platform by executing our two-step method to make crystals that exhibit bulk optical properties using particles with diameters of 600~nm (\figref{fig:crystals}c), 430~nm (\figref{fig:crystals}d), and 250~nm~(\figref{fig:crystals}e).
For each particle size, we successfully self-assemble similarly large single crystals that rival the sizes of crystals formed from nanometer-scale particles (see \figref{fig:crystals}b for comparison). Furthermore, we find that each of these particle sizes assembles into a different crystal structure, indicating that our method can be applied to the synthesis of different crystal symmetries. The 600-nm-diameter particles assemble into a crystal that is nearly isostructural to CuAu-I; the 250-nm-diameter particles assemble into a crystal that is isostructural to CsCl; and the 430-nm-diameter particles assemble into a tetragonal crystal structure that is intermediate between the CsCl and CuAu-I. Because the precise crystal structures, are not the primary focus of this article, we have placed a detailed discussion of their characterization in the SI.

Owing to size of the constituent building blocks, together with the  size and quality of the crystalline assemblies, our crystals show pronounced photonic properties. 
For example, the 250-nm-diameter-particle crystals exhibit a prominent stop band for frequencies corresponding to red light and therefore exhibit structural coloration in reflected light (\figref{fig:crystals}e). The
400-nm-diameter-particle crystals also exhibit coloration in reflection (\figref{fig:crystals}d). We hypothesize that the green-yellow structural color of the 400-nm-diameter-particle crystals arises from second-order diffraction, which explains why the apparent color is shorter wavelength even though the particle size is larger than that of the 250-nm-diameter particles (see SI Section III F for details).

We stress that the specific photonic properties of our crystals are not the most exciting result, as large colloidal crystals exhibiting structural coloration have been made by numerous other methods~\cite{colvin2001opals,zhang2009self,hou2018patterned,jiang2021centimetre}.
Rather, the notable achievement here is that our method enables the robust, near-to-equilibrium assembly of macroscopic single crystals from \emph{DNA-coated colloids} that can grow to macroscopic sizes visible to the naked eye (\figref{fig:crystals}b and Fig.~\SIfigCellphone).
The importance of demonstrating the assembly of macroscopic crystal domains from DNA-coated colloids is that many of the other reported methods for making large-scale colloidal crystals lack the programmability of DNA, which is essential for creating complex crystalline lattices, like cubic diamond~\cite{he2020colloidal}, for advanced photonic bandgap materials.
We expect that our two-step platform is applicable to any Brownian suspension of DNA-coated colloids (i.e., particles with diameters less than roughly two micrometers).
Furthermore, because the interactions that drive nucleation and growth are due entirely to DNA hybridization, our method should also apply to a wide range of colloidal particle compositions, including polymers, metals, metal oxides, semiconductors, and magnetic materials~\cite{macfarlane2011nanoparticle,Wang2015,wang2015synthetic,Pine2015ChemMater,zhang2013general}.
We similarly anticipate that our theoretical model of droplet-confined nucleation and growth under a time-dependent protocol could be extended to other particle sizes and crystal structures with minimal modifications.

\section{Discussion}

In summary, we have developed a new platform to create macroscopic single crystals from DNA-coated colloids by decoupling nucleation and growth.
Our approach solves a number of longstanding challenges associated with isothermal nucleation in bulk solution and the resulting heterogeneous distribution of relatively small self-assembled crystals.
We first showed that our method for assembling monodisperse crystalline seeds in droplets is theoretically near optimal, despite fundamental physical constraints on the size of seeds that can be formed in this manner.
We then demonstrated a practical and reliable strategy for controlling secondary growth in bulk solution, which preserves the narrow size distribution due to the deterministic nature of diffusion-limited growth.
Although seeded growth of colloidal crystals has been explored before, previous attempts have focused primarily on growing thin shells~\cite{seo2020modulating}, growing crystals from two-dimensional seeds \cite{ganapathy2010direct,C0SM01219J,lewis2020single}, or using seeds that do not match the thermodynamically favored crystal structure \cite{Allahyarov2015NComm}.
By contrast, our method uses three-dimensional seeds that are isostructural to the crystal that we wish to grow.
The end result is a monodisperse distribution of single crystals whose sizes are precisely controlled by the duration of the seeded growth process.

By growing single crystals more than two orders of magnitude larger than was previously possible using optical-scale DNA-coated colloids, our protocol accomplishes the longstanding goal of assembling DNA-programmed materials with user-specified photonic properties. 
We fully expect that our method could be extended to make single crystals with other crystallographic symmetries~\cite{macfarlane2011nanoparticle,Tkachenko10269}, including ones that can only be synthesized by DNA-programmed assembly such as colloidal diamond~\cite{he2020colloidal}, by changing the constituent particles.
Furthermore, incorporating additional strategies for processing same-size single-crystalline assemblies, as was recently demonstrated using nanoscale building blocks~\cite{santos2021macroscopic}, could open up additional possibilities for hierarchical materials engineering.
By providing robust routes to assemble bespoke metamaterials, the advances developed herein promise to bring DNA-programmed colloids out of the lab and into practical use.

\begin{acknowledgments}
We thank Larry Luster for initial conversations about seeded growth. This work was supported by the National Science Foundation (DMR-1710112).
\end{acknowledgments}


\bibliography{maintext}

\end{document}


\onecolumngrid

\section*{Supplementary Information for\\``Macroscopic DNA-programmed photonic crystals via seeded growth''}

\section{\label{sec:Materials}Materials and Methods}
\noindent \textbf{Synthesizing DNA-coated colloids} \\
Colloidal particles are functionalized with DNA using a combination of strain-promoted click chemistry and physical grafting, following a modified version of the methods described by Pine and co-workers~\cite{Pine2015ChemMater}. In brief, polystyrene-block-poly(ethylene oxide) (PS-b-PEO) copolymers are functionalized with an azide group, the azide-modified PS-b-PEO is adsorbed onto the surface of polystyrene colloidal particles, and then DBCO-modified DNA is attached to the PS-b-PEO via click chemistry. After the reaction, the particles are washed five times in 1xTE via centrifugation and resuspension and stored at 1\% (v/v). A detailed protocol is provided in the Supplementary Information of Reference~\cite{Hensleye2114050118}. 

We study crystallization of a binary mixture of same-sized DNA-coated colloids. One particle species is coated with sequence A: 5’-(T)51-GAGTTGCGGTAGAC-3’; the other particle species is coated with sequence B: 5’-(T)51-AATGCCTGTCTACC-3’. Both DNA sequences are obtained from Integrated DNA Technologies (IDT) and are purified by high-performance liquid chromatography (HPLC). All crystallization experiments are performed in 1x tris-EDTA buffer (1xTE) with 500~mM NaCl. In experiment, these sequences yield a melting temperature of roughly 52 degrees Celsius for the 600-nm-diameter particles, which is consistent with the predicted melting temperature from an experimentally-validated mean-field model of DNA-mediated colloidal interactions~\cite{rogers2011direct,rogers2020mean}. More specifically, taking $\Delta H=-56.7$~kcal/mol and $\Delta S=162.5$~cal/mol-K from the Nearest Neighbor model and a surface density of 5,000 DNA per particle, we predict a melting temperature of 54 degrees Celsius, assuming that melting occurs when the binding free energy is 6~$k_B T$, in agreement with our experimental observations. We note that this grafting density is consistent with our previous estimates~\cite{fang2020two} and with  measured DNA surface densities that lead to crystallization~\cite{Wang2015} from a related click-chemistry method. \\

\noindent \textbf{Fabricating the microfluidic device} \\
Microfluidic drop-makers are fabricated via standard photolithographic techniques. A glob of SU8 (SU-8 2075, or SU-8 3010 MicroChem) roughly the size of a quarter is poured onto a silicon wafer (3-76-024-V-B, Silicon Materials Inc.). The wafer is then spun at 500~rpm with a spin coater at a ramp rate of 100~rpm/sec for 5 seconds, and then to between 1000~rpm and 3000~rpm at a ramp rate of 300~rpm/sec for 60 seconds, the specifics of which will lead to a device thickness between 20~$\mu$m and 80~$\mu$m. Next, the wafer is placed onto a 65~$^\circ$C hot plate for 3 minutes and then a 95~$^\circ$C hot plate for 5 minutes. A photomask (Output City) with the pattern of our microfluidic device is placed on top of the wafer, which is then moved to a Manual Mask Aligner System (ABM-USA) and exposed to UV light for 46 seconds for a total of 160~mJ. The mask is removed and the wafer is washed with isopropanol and propylene glycol methyl ether to remove the undeveloped photoresist. The wafer is then dried with an air brush and placed on a 65~$^\circ$C hot plate for 3 minutes and a 95~$^\circ$C hot plate for 20 minutes. Next, the wafer is placed in a glass Petri dish with PGME and shaken back and forth for 10 minutes to remove any photoresist. Finally, the wafer is sprayed with isopropanol and dried with an air brush.

The master is a negative of the actual device and acts as a mold. 30~g of polydimethylsiloxane (PDMS) and 3~g crosslinker (1673921, Dow Chemical Company) is mixed using a Thinky AR-250 planetary centrifugal mixer for 3 minutes. A plastic Petri dish is lined with aluminum foil and the microfluidic-device master is placed face up in the dish. The mixed PDMS is then poured onto the master and placed in a vacuum desiccator for 30 minutes to remove any bubbles from the PDMS mixture. The dish is placed in a 70~$^\circ$C oven overnight. The wafer is removed from the dish, the foil is peeled off, and a hobby knife is used to cut away the excess PDMS and separate it fully from the master. A coring tool (69039-07, Electron Microscopy Sciences) is then used to punch holes into all the device inlets and outlets. A glass slide (2947-75X50, Corning) and the PDMS chip are placed into an oxygen plasma cleaner (Zepto, Diener electronic) for 45 seconds. The PDMS chip is then laid down onto the glass slide and held with uniform pressure for 30 seconds, permanently bonding them together. Further details are provided in the Supplementary Information of our previous paper\cite{Hensleye2114050118}.\\

\noindent \textbf{Droplet making} \\
Syringe pumps are used to operate the microfluidic device to produce monodisperse droplets containing a colloidal suspension. The channels of the microfluidic device are made hydrophobic by flushing them with Aquapel (B004NFW5EC, Amazon), leaving it for 30 seconds, and then flushing them again with air to remove the Aquapel. The channels are then flushed with HFE-7500 oil (3M) and air again. Flow rates are controlled independently by three syringe pumps (98-2662, Harvard Apparatus) connected to the device via tubing (06417-11, Cole-Palmer) that is slightly larger in diameter than the holes to ensure a snug fit. HFE-7500 with 2.5\% RAN fluorosurfactant (008-FluoroSurfactant-5wtH-20G, RAN Biotechnologies) is fed into the oil inlet, 1~M NaCl in 1xTE buffer is fed into one aqueous inlet, and DNA-coated particles suspended at twice the desired volume fraction in 1xTE are fed into the other aqueous inlet. The particles are created in small quantities so we cannot load them directly into the syringe. Instead they are loaded into the tube by using a reverse flow rate and never enter the syringe body. A couple tube centimeters of air are left on either side of the particle solution to ensure that the suspension does not mix with the carrier fluid due to Taylor dispersion. The flow rates of the oil and aqueous phases depend on the desired droplet size and the thickness of the microfluidic device being used and are generally between 400~$\mu$l/hr and 1000~$\mu$l/hr. The droplets are deposited from the outlet tube directly into a 0.2~ml Eppendorf tube. As much as 10~ul of HFE-7500 with 2.5\% RAN is added to the tube if the ratio of oil to aqueous flowrates was lower than 1:1. A very small amount of droplets are loaded directly into a glass capillary and the droplet size is verified via brightfield microscopy. \\

\noindent \textbf{Droplet temperature ramp experiment} \\
Eppendorf tubes with particle-filled droplets are placed in the central wells of a single module C1000 Touch Thermo Cycler (Bio-Rad). An Eppendorf with a thermistor and thermal paste is placed in a well next to the sample to log the sample temperature. A ramp protocol is used that comprises of 30 minutes of melting at 56$^{\circ} C$ followed by a drop to the temperature at which the ramping protocol begins. The ramping protocol involves dropping the temperature in 0.1$^{\circ} C$ increments and holding for a specific interval defined by the quoted ramp rate of the experiment. For instance, for a ramp rate of 0.025$^{\circ} C/hr$ a 0.1$^{\circ} C$ drop every 4~hours would be used. The ramp continues for 40 steps covering a range of 4$^{\circ} C$. The starting temperature is decided by placing a small quantity of particles in buffer in an Eppendorf and observing whether the particles aggregate and sink over the course of 30 minutes. Once this transition temperature is found the starting temperature is set 1.5$^{\circ} C$ degrees above it. \\

\noindent \textbf{Fabricating sample chambers} \\
Sample chambers to observe the presence of crystals in the droplets are comprised of a rectangular capillary filled with the microfluidic emulsion that is glued to a glass coverslip. A 200-$\mu$m tall, 2-mm-wide glass rectangular capillary (5012, VitroCom) is cut to 3~cm in length with a glass scoring pen and held suspended in place with a pair of clamping tweezers. Approximately 2--3~$\mu$l of the droplet emulsion is transferred into the capillary via a micropipette that has been snipped at the tip to have a wider inlet. HFE 7500 with 2.5$\%$ RAN is used to fill the rest of the volume. The capillary is then placed on a glass slide and sealed with two-part epoxy (BSI-202, Bob Smith Industries). The sample is cured for roughly 30~minutes. Care is taken to ensure that no air bubbles are present in the tube during sealing. Ultimately, these slides are placed into an acrylic holder on the microscope stage that positions the capillary down facing the objective of our microscope.\\

\noindent \textbf{Brightfield imaging and counting crystals} \\
Brightfield microscope images are obtained using a Nikon Ti2 microscope with a 10x-magnification, 0.45 NA objective (MRD00105, Nikon), a 1.5x-magnification tube lens, and a Pixelink M12 Monochrome camera (M12M-CYL, Pixelink) connected to a desktop computer. The focus is set such that a majority of the presented face of the crystals are in good focus. To maintain focus across multiple fields of view we use the Nikon Perfect Focus System.\\

\noindent \textbf{Polarized light imaging} \\
Polarized reflection and transmission microscopy images were taken either on an Olympus BX51 microscope with an incandescent lamp or a Nikon Ti2 microscope with a white LED illuminator. Crossed polarizers are installed in each case and the images are taken with a color CMOS Camera (CS895CU, Thorlabs). The analyzer is always aligned perpendicular to the polarizer for maximum contrast. We image some crystals at different angles relative to the polarizer. For these images, the angle of the polarized light is shifted by 5$^{\circ}$ between each image by rotating both the polarizer and the analyzer in tandem. It is only necessary to image $90^\circ$ of rotation as the other quadrants are symmetrical. To obtain polarized light images of crystals while they were growing, the sapphire window on the Peltier unit had to be replaced by a glass one as the sapphire affected the polarization of the incident light.\\

\noindent \textbf{Seeded growth experiments} \\
Droplets filled with crystals are mixed in an Eppendorf tube with droplets filled with particles that have half the DNA density as compared to the seed particles. To get an acceptably low density of seeds in the final experiment, 1~$\mu$l of droplets with seed crystals is added to 4~$\mu$l of droplets with weak particles. Then 1~$\mu$l of this mixture is added to 4~$\mu$l of droplets and so on for a total of three dilutions. Finally, a fourth dilution adds 1~$\mu$l of this diluted mixture to 9~$\mu$l of droplets containing weak particles. This solution is then loaded into a capillary  until the capillary is completely full, sealed using UV-glue (Norland Products, NOA68), and then cured for at least 10 minutes under a mercury vapor UV-lamp. 

The sample is brought to a microscope and is heated to a temperature at which the weak particles disassociate but the seeds remain intact. The sample is then quickly brought to an analytical balance with an attached ionizer (Mettler Toledo XSE104) and is gently moved back and forth across the ionizer aperture for 30 seconds. The ionizer breaks the emulsion, combining the particles with the seeds. The sample is put on the microscope again and is heated using a thermoelecric cooler to melt the weak particles until the combined system is in equilibrium. A detailed description of the sample heater is provided in the Supplementary Information of Reference~\cite{Hensleye2114050118}. 

The field of view is centered on a region with the desired density of seeds and a time-lapse video of growth is recorded. We acquire one picture every five minutes. To maintain focus over the duration of the experiment we use the Nikon Perfect Focus System. On each image, the projected area of  a chosen reference crystal is measured using image analysis routines written in Matlab.  The system starts at a temperature at which crystal growth does not occur and the temperature is automatically lowered in $0.05^\circ${C} steps until the reference crystal begins to grow. 

We notice a very slow, apparent `weakening' of the DNA-mediated interactions during the seeded growth experiments, which we compensate for by gradually lowering the temperature. The evidence of the apparent weakening of the interactions is that the assembled crystals begin to melt after a few hours of growth at constant temperature.  We attribute this effect to an interaction between the DNA-coated colloidal particles and the residual oil/surfactant that is left over from the broken emulsion, which we hypothesize interferes with the surface-grafted DNA molecules, thereby lowering the melting temperature of the suspension over time. We highlight that the weakening of the interactions is inconsistent with simply a reduction in the supersaturation due to the depletion of monomers, because it is physically impossible that a reduction in the supersaturation would lead to crystal melting. At the very most, the crystal and gas would reach equilibrium coexistence and growth would cease. 

\begin{figure}[h]
  \centering \includegraphics{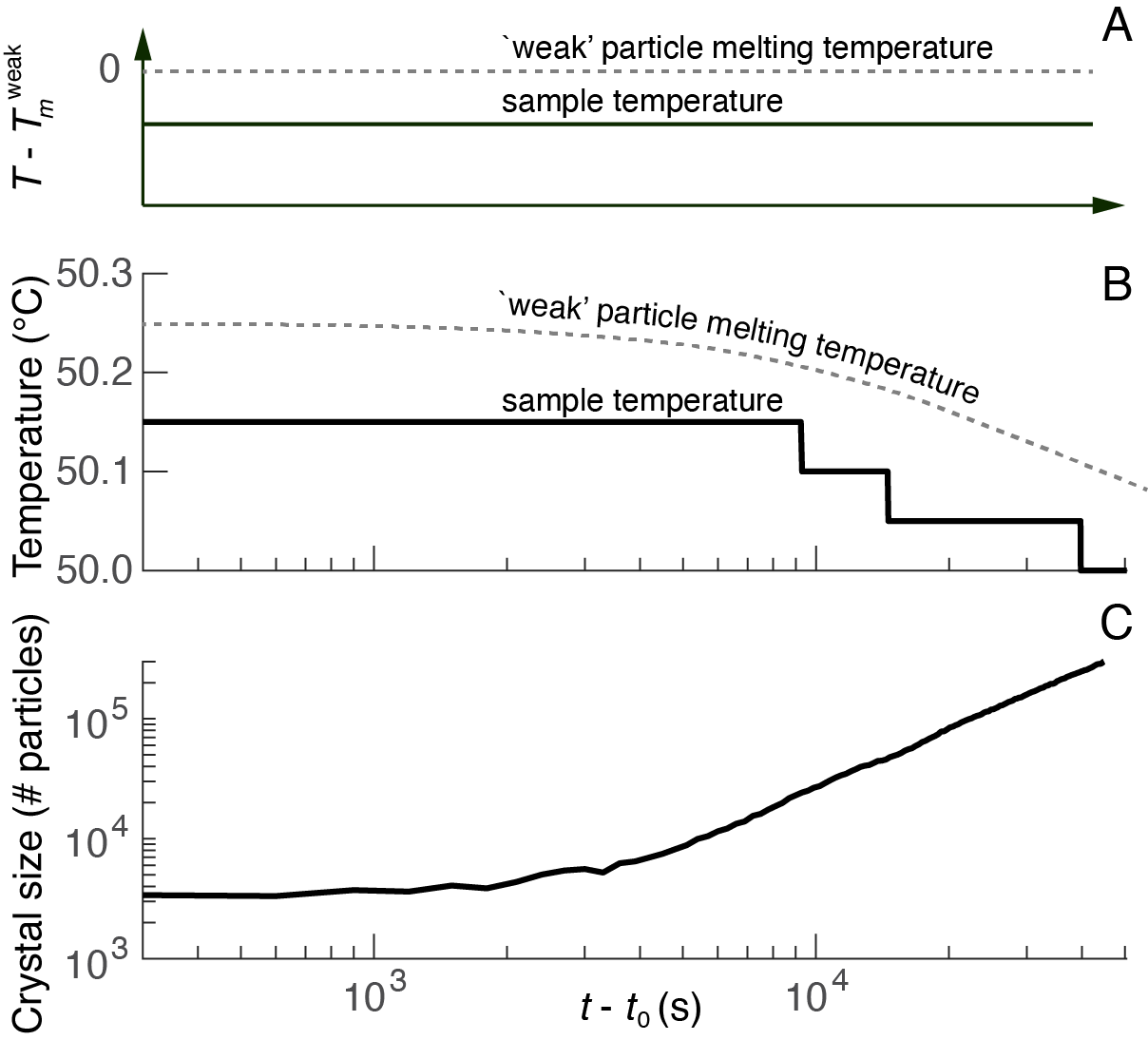}\vskip-1ex
  \caption{\label{fig:supersaturation}
  Plots showing the sample temperature relative to the weak particle melting temperature (A), the weak particle melting temperature and the sample temperature (B), and the size of the reference crystal (C) as a function of time. The sample temperature is decreased three times across the experiment keep the difference between the sample temperature and the melting temperature constant, thereby preventing the crystal from shrinking in size.
  }
\end{figure}

To offset the `weakening' of the DNA-mediated interactions over time, we track the  area of the same reference crystal continuously throughout the experiment and decrease the temperature if we see that the crystal begins to shrink. More specifically, after growth begins, if the area of this crystal decreases between two successive frames (i.e., after 10 minutes), a counter is incremented. Once this counter reaches two, the temperature is lowered by $0.05^\circ${C}.  An example of the temperature trace over an entire experiment is shown in Fig.~\ref{fig:supersaturation}, where the temperature is lowered by $0.15^\circ$C over the course of roughly 14~hours. We stress that reducing the temperature with time is simply to maintain a constant supersaturation and not to accelerate growth.

\section{Data analysis}

\begin{figure}[h]
  \centering \includegraphics[width=0.6\textwidth]{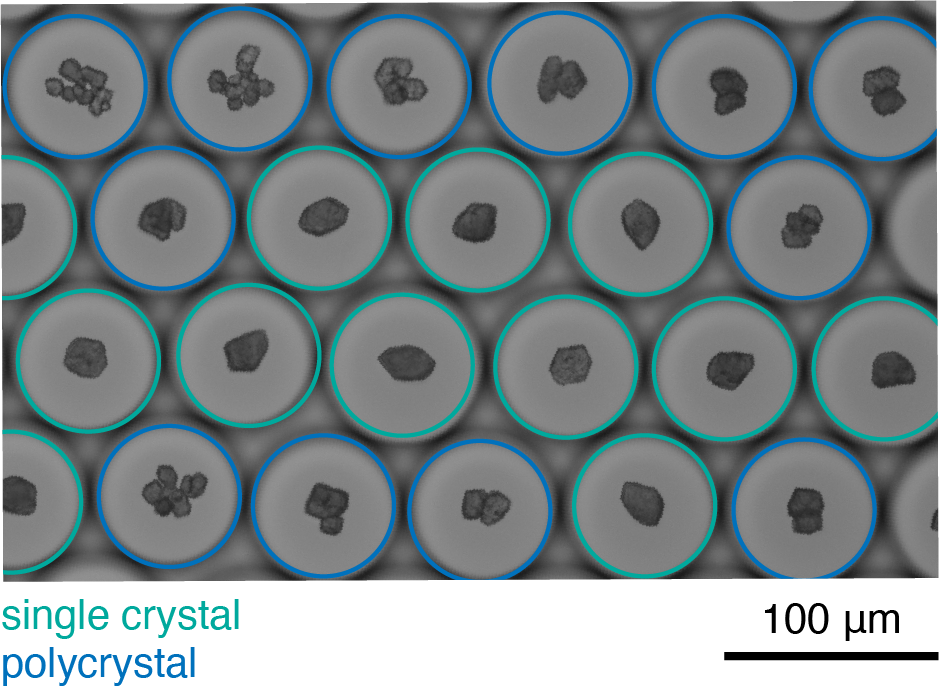}\vskip-1ex
  \caption{\label{fig:P1Color}
    An example image of crystals from a droplet temperature ramp experiment. Some droplets have single crystals (green), others have polycrystals (blue).}\end{figure}

\subsection{Determining the single crystal fraction}
Single crystals are identified manually to compute the single crystal fraction. In brief, an image of an entire camera field of view is loaded into a standard computer paint program at a magnification that is sufficient to make out the details in the individual crystals (Fig.~\ref{fig:P1Color}. Single crystals or polycrystals are determined by eye by looking either for the presence of grain boundaries or for an irregularly shaped crystal as opposed to a well faceted crystal. Droplets that appear to contain a single crystal are marked with a red dot and droplets that appear to contain a polycrystal are marked with a blue dot. These images are saved and then analyzed via a MATLAB script that tallies the number of red dots, $N_{\text{r}}$, and blue dots, $N_{\text{b}}$, and then calculates the proportion of single crystals as $p_{\text{1x}} = N_{\text{r}}/(N_{\text{r}} + N_{\text{b}})$. Droplets that are out of focus or appear to have an anomalously low or high numbers of particles are excluded from the analysis.

\subsection{Determining the crystal size}
\noindent\emph{Seeded growth.} We determine the number of particles per crystal for our seeded growth experiments using an empirical calibration curve based on the projected area of the single crystals. For each of the distinct droplet volumes used in the temperature-ramp experiments (Fig.~2B-C in the main text), we measure the crystal area of a large population of single crystals. For each droplet volume, we know the total number of particles inside the droplet since we know the total particle volume fraction: $N_{\text{drop}} = V_{\text{drop}}\rho_0$, where $\rho_0$ is the known concentration of particles used in microfluidic drop making and $V_{\text{drop}}$ is the measured volume of a droplet in the given experiment. Because all of the particles are incorporated into the crystal phase by the end of the temperature-ramp experiment, we know that $N_{\text{crystal}} = N_{\text{drop}}$. 

\begin{figure}[h]
  \centering \includegraphics[width=0.6\textwidth]{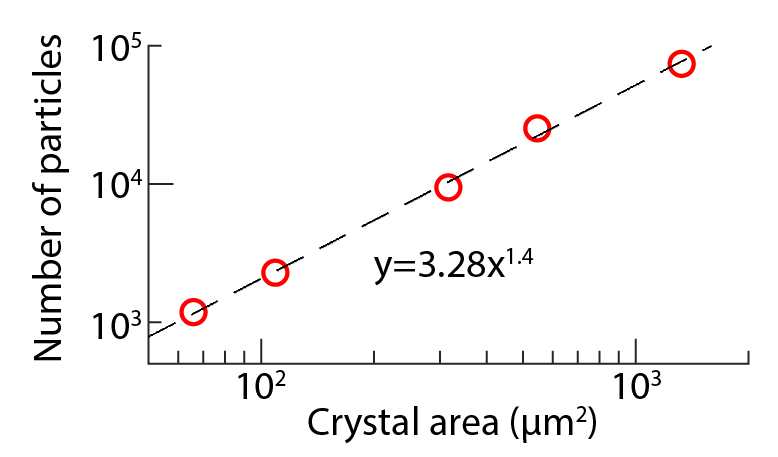}\vskip-1ex
  \caption{\label{fig:area_number}
    The number of particles in a crystal is extrapolated from the projected area of the crystal via the empirical expression $N_{\text{crystal}} = 3.28 A_{\text{crystal}}^{1.4}$.}
\end{figure}

We plot the mean crystal area with respect to the number of particles inside the droplet and find that it follows a power law (Fig.~\ref{fig:area_number}). A fit of a power law to our data shows that the number of $600$-nm-diameter particles per crystal follows the scaling relation $N_{\text{crystal}} = 3.28 A_{\text{\text{crystal}}}^{1.4}$, where $A_{\text{crystal}}$ is the crystal area in $\mu m^2$. We then use this expression to infer the number of particles per crystal for our seeded growth experiments (see Fig.~3 in the main text). Note that the expected power-law dependence would be $N_{\text{crystal}} \propto A_{\text{crystal}}^{3/2}$. We hypothesize that the small difference is due to the fact that the crystals are slightly more dense than water and thus settle to the bottom of the sample chamber once they are sufficiently large.\\

\noindent \emph{Comparing to literature.} Because we cannot use the same calibration curve to estimate the volumes of crystals made from colloidal particles of differing diameters reported in the literature, we estimate the crystal sizes using a much more approximate method. We estimate the crystal sizes in Fig.~4 by taking the cube of the linear dimension of the crystal $s$ relative to the particle diameter $d$: $N\approx (d/s)^3$. We expect that this estimate is an upper bound of the crystal volume, as evidenced by our calibration method described above.

\subsection{Determining the start of growth for the seeded growth experiments}
The seeded growth experiments are performed by starting at a temperature above the melting temperature of the `weak' particles, but below the melting temperature of the seeds, and then lowering the temperature slowly until the crystals start growing. The start time of crystal growth $t_0$ is therefore determined by the frame at which the reference crystal first reaches a value $10\%$ larger than its initial size. 

\section{Solving the crystal structures and crystal habits}

In the main text we show the crystallization and seeded growth of single-crystalline assemblies of three different binary mixtures of same-sized particles with diameters of 600-nm, 430-nm, and 250-nm. The following section describes the quantitative pipeline that we developed to characterize their 3D crystal structures, including both the lattice structure and the crystal habit. 

\subsection{An overview of the binary body-centered tetragonal crystal structure}
\label{sec:BCT}

\begin{figure}[h]
  \centering \includegraphics[width=\textwidth]{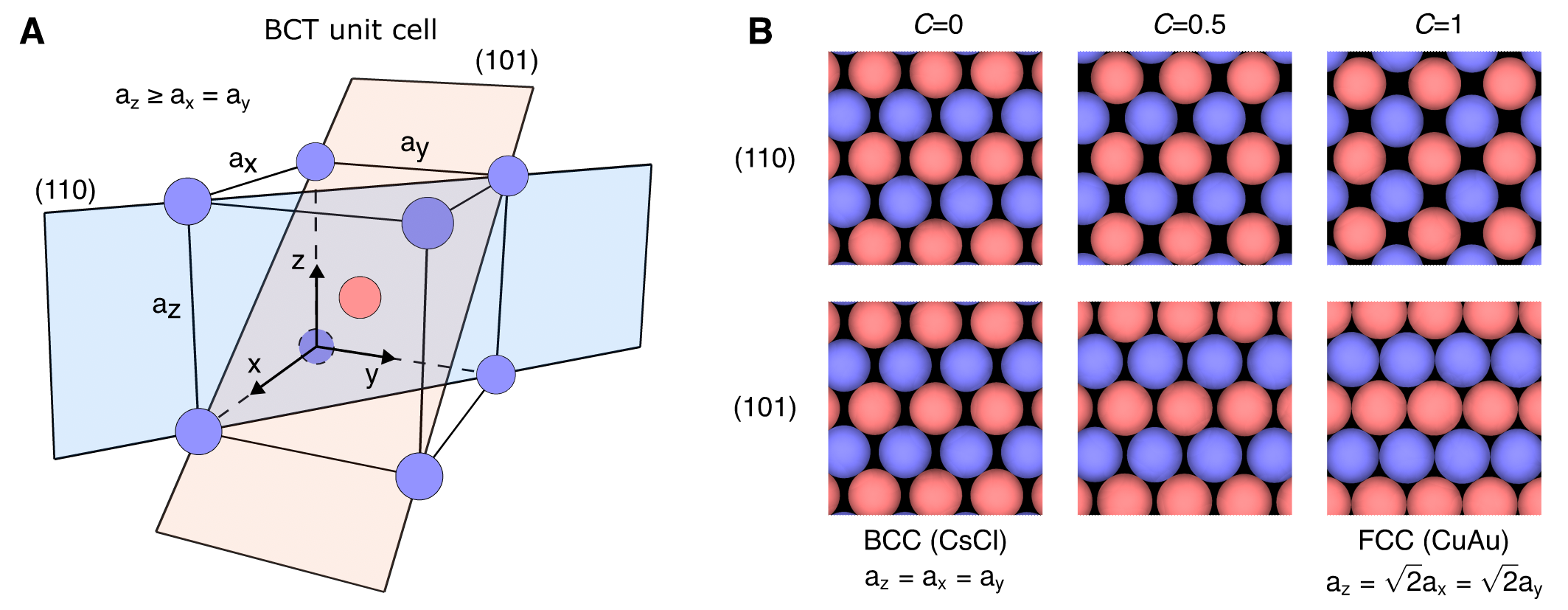}
  \caption{(A) An illustration of a BCT unit cell. Note that with this definition of the BCT structure, we consider the vertical lattice spacing, $a_z$, to be greater than or equal to the other lattice spacings, $a_x$ or $a_y$. The facets with the lowest surface energies for this structure are the (110) and (101) planes, shown by the blue and red planes, respectively. (B) Images of the (110) and (101) facets for different BCT configurations of $C$=0,0.5,1.0. Note that when $C=0$, the BCT lattice is the same as BCC-CsCl, and when $C=1$, the BCT lattice is the same as FCC-CuAu.}
  \label{SI:fig-BCTunitcell}
\end{figure}

All three crystal structures that we find can be classified within a binary version of the body-centered tetragonal (BCT) crystal structure. The binary BCT structure is characterized by having primitive vectors of $a_x\hat{x}$, $a_y\hat{y}$, and $a_z\hat{z}$, with the restriction that $a_x=a_y$ and the two sublattices of the binary mixture displaced by a vector $(a_x\hat{x}+ a_y\hat{y}+a_z\hat{z})/2$, as shown in Fig.~\ref{SI:fig-BCTunitcell}A. 

The class of binary BCT crystal structures has within it crystals that are isostructural to BCC-CsCl and FCC-CuAu. When $a_z=a_x=a_y$, we recover a binary simple cubic lattice, isostructual to CsCl, and when $a_z = \sqrt{2}a_x=\sqrt{2}a_y$, the lattice is isostructural to the FCC-CuAu lattice. To clearly see these two limits we show the (110) and (101) planes of the binary BCT lattice in Fig.~\ref{SI:fig-BCTunitcell}B on the left and right sides. Note that the (110) and (101) binary BCT planes are the same as the CsCl (110) and (101) planes (left), while in the other limit the binary BCT (110) is the same as the CuAu (100) plane and the binary BCT (101) plane is the same as the CuAu (111) plane. We choose only to show the (110) and (101) planes of the binary BCT structure since these are the lowest-surface-energy facets and are the mostly like to be present in the facets of crystals that form. This note will be described in more detail in the \secref{sec:habits}.

To characterize the BCT structures that lie between CsCl and CuAu, we introduce a linear transformation for the magnitudes of $a_x$ and $a_z$,
\begin{eqnarray}
a_x=a_y=\frac{2}{\sqrt{3}}a_0(1-C) + a_0 C \nonumber \\
a_z = \frac{2}{\sqrt{3}}a_0(1-C) + \sqrt{2}a_0 C, \nonumber
\end{eqnarray}
where $a_0$ is the spacing between the two particle types and $C$ is a scalar. When $C=0$, we recover  CsCl, and when $C=1$, we recover  CuAu. As an example, we show the (110) and (101) planes for $C=0.5$ in the middle column of Fig.~\ref{SI:fig-BCTunitcell}B. We note that if $a_0$ remains fixed, this transformation preserves each particle having eight nearest-neighbor contacts of the other particle type.

\subsection{A quantitative method to characterize the 3D crystal structure}
\label{sec:pipeline}

We develop a quantitative pipeline based on laser-scanning confocal microscopy, image analysis, and crystallography to determine the 3D crystal structures of the three different crystal types that we explore in experiment. The pipeline is as follows: (i) collect images of the crystal facets that have sedimented during growth; (ii) find particle positions for both A and B particles using image analysis and compute their radial distribution functions, $G_{AB}(r)$ and $G_{AA}(r)$ (note that we only use $G_\mathrm{AA}(r)$ when one of the particles cannot be resolved); (iii) generate a look-up table of model $\Tilde{G}(r)$ for potential crystalline facets (from here on $\Tilde{G}$ will denote model functions); and (iv) compare the experimental $G(r)$ against all model $\Tilde{G}(r)$ to find the closest match. As a complimentary technique, we also take a scan orthogonal to the exposed facet to look at the crystalline order in the third dimension. This complementary method is described in \secref{sec:internal}.

\begin{figure}[h]
  \centering \includegraphics[width=\textwidth]{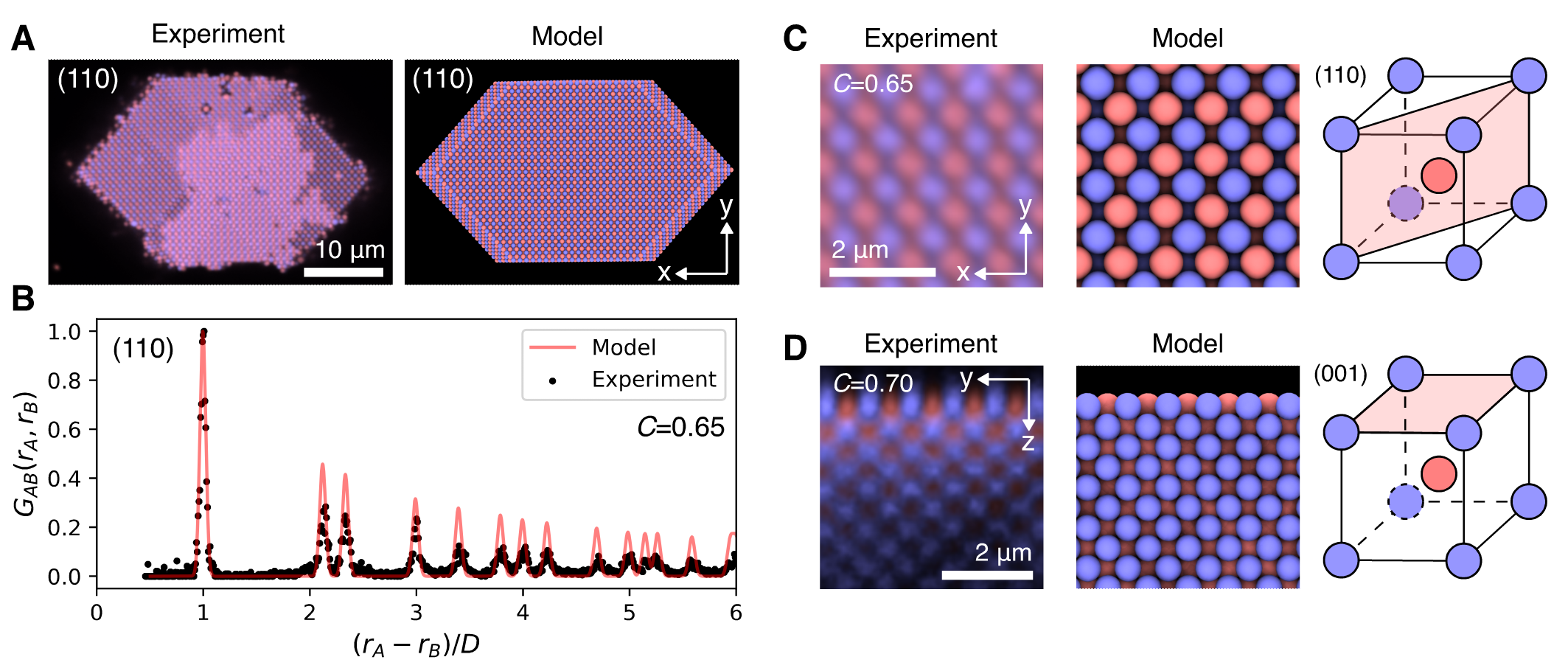}
  \caption{\textbf{600-nm-diameter particle crystal characterization.} (A) A confocal image (left) of the presented facet of one crystal formed from a binary mixture of 600-nm-diameter particles as compared to the best-fit model (right). (B) The pair correlation function for unlike particle species showing both experimental data (points) from the image in (A) and a best-fit model from a BCT (110) facet with $C$=0.65. (C) Zoom in of the facet of the image in (A) along with a zoom in of the corresponding model BCT (110) facet. The facet is denoted by the red plane in the unit cell illustration. (D) A Z-slice confocal image of a different crystal with BCT $C$=0.70 with a corresponding model slice. The facet is denoted by the red pane in the unit cell illustration. Note that the experimental image comes from a projection of four layers of the crystal as described in \secref{sec:internal}.}
  \label{SI:fig-600nmcrystal}
\end{figure}

We image the crystals with a Leica SP8 laser-scanning confocal microscope. Since our crystals are composed of two particle species, we independently dye each particle type, one with Pyrromethane 546 and one with Nile Red. We then take a two-color acquisition to capture the particle locations and types for a given crystallographic plane, as shown in Fig.~\ref{SI:fig-600nmcrystal}A. The crystal facet images are then analyzed to find particle positions~\cite{crocker1996particleid}. In brief, the images are split into separate channels corresponding to the two particle types in the crystal; they are convolved with a Gaussian to reduce the noise; a local thresholding algorithm is used to isolate the center of particles and to compensate for any large-scale variations in image intensity; and lastly we compute the centroids to find the positions of all the particles in the facet. The positions of the particles are then used to calculate the experimental radial distribution function, $G_(r)$~\cite{chandler}.

To create the reference $\Tilde{G}(r)$ data, we identify relevant facets of proposed crystal structures that may show up in experiment. We then generate position data of A and B particles for those facets given a proposed crystal structure, as shown for an example model facet in Fig.~\ref{SI:fig-600nmcrystal}A. In our experiments, we primarily see the (110) and (101) planes of body-centered tetragonal (BCT) structures described in \secref{sec:habits}. Once the particle positions are defined, we calculate corresponding $G(r)$ data sets. In our lookup table we included the (100), (110), (111), and (210) facets of binary BCC (CsCl) and FCC (FCC-CuAu) structures, as well as the (110), (101) facets of BCT structures with $C$-values ranging from 0 to 1 in steps of 0.05. Other facets, such as the (100) or (111) facets of BCT, were excluded because we do not observe facets that contain only a single particle type.

Next, we compare the experimental and model $G(r)$ data to identify the best match.  In order to make this comparison, we need to introduce noise into our model data that is comparable to the noise in our experimental data. We fit the $G(r)$ peak that is closest to $(r_A - r_B)/D=1$ (2 for $G_\mathrm{AA}(r)$) with a Gaussian and extract the mean and standard deviation. The mean is used to get the true lattice spacing of the crystal and to rescale the experimental $G(r)$ to have a peak at $(r_A - r_B)/D$ at 1 (2 for $G_\mathrm{AA}(r)$). We then convolve the model data with a Gaussian that has the same standard deviation as the experiment. Both sets of data are then normalized to set the amplitude of the first peak to one. This normalization sets a reference point from which we can compare various model facets to our experiment. For a given model $\Tilde{G}(r)$, we compute the sum of the squared residuals between the model and the experimental data, $\sum_i( G_(r_i)-\Tilde{G}(r_i))^2$. We take the model with the smallest sum of the squared residuals as the best fit crystal structure to the experimental image. An example best fit is shown for 600-nm-diameter particles in Fig.~\ref{SI:fig-600nmcrystal}B. 

A similar analysis can also be performed for internal crystal planes, as described in \secref{sec:habits}. By taking a vertical scan through the crystals, we can observe the internal order and subsequently fit particle positions to find a best fit crystal structure. An example of such a plane for the 600-nm-diameter particles in shown in Fig.~\ref{SI:fig-600nmcrystal}D with a corresponding model plane next to it.

\begin{figure}[h]
  \centering \includegraphics[width=\textwidth]{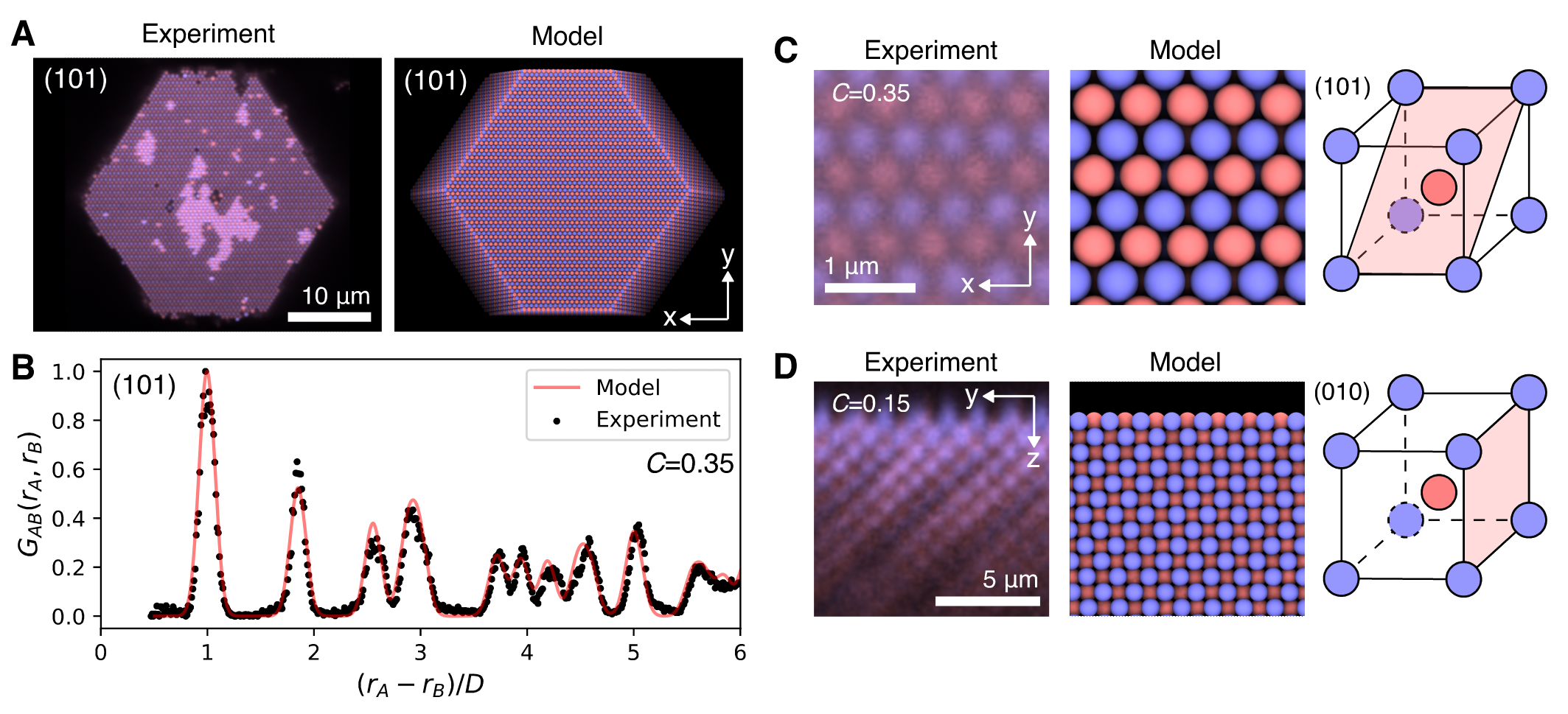}
  \caption{\textbf{430-nm-diameter particle crystal characterization.} (A) A confocal image (left) of the presented facet of one crystal formed from a binary mixture of 430-nm-diameter particles as compared to the best-fit model crystal (right). (B) The pair correlation function for unlike particles showing both experimental data (points) from the image in (A) and a best fit model from a BCT (101) facet with $C$=0.35. (C) Zoom in of the facet of the image in (A) along with a zoom in of the corresponding model BCT (110) facet. The facet is denoted by the red plane in the unit cell illustration. (D) A Z-slice confocal image of a different crystal with BCT $C$=0.15 with a corresponding model slice. The facet is denoted by the red plane in the unit cell illustration. Note that the experimental image comes from a projection of four layers of the crystal as described in \secref{sec:internal}.}
  \label{SI:fig-430nmcrystal}
\end{figure}

\begin{figure}[h]
  \centering \includegraphics[width=\textwidth]{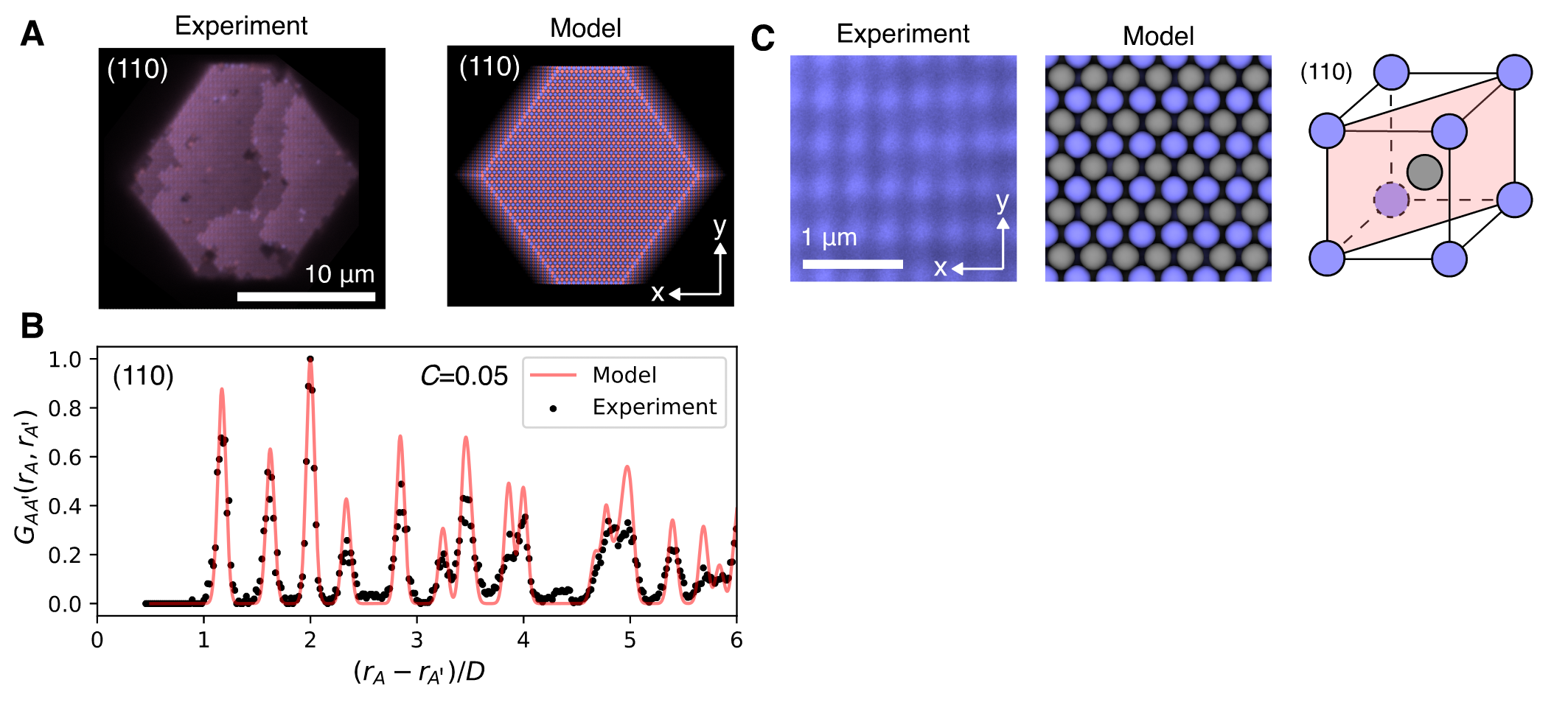}
  \caption{\textbf{250-nm-diameter particle crystal characterization.} (A) A confocal image (left) of the presented facet of one crystal formed from a binary mixture of 250-nm-diameter particles as compared to the best-fit model crystal (right). (B) The pair correlation function for like particles showing both experimental data (points) from the image in (A) and a best fit model from a BCT (110) facet with c=0.05. Because we were unable to resolve the red particles at this particle size, they were not used for the analysis of the lattice structure. (C) Zoom in of the blue particles on the facet of the image in (A) along with a zoom in of the corresponding model BCT (110) facet. The facet is denoted by the red plane in the unit cell illustration. Note that the experimental image comes from a projection of four layers of the crystal as described in \secref{sec:internal}.}
  \label{SI:fig-250nmcrystal}
\end{figure}

Example results of this pipeline for crystals composed of 430-nm and 250-nm particles are shown in Figs.~\ref{SI:fig-430nmcrystal} and \ref{SI:fig-250nmcrystal}. We note that due to the size of the 250-nm particles, we were unable to resolve the Nile Red-dyed particles because they are diffraction-limited, so only the Pyrromethane 546-dyed particles are used to compute $G_{AA}(r)$.

\subsection{Imaging the internal crystal structure}
\label{sec:internal}

\begin{figure}[h]
  \centering \includegraphics[width=\textwidth]{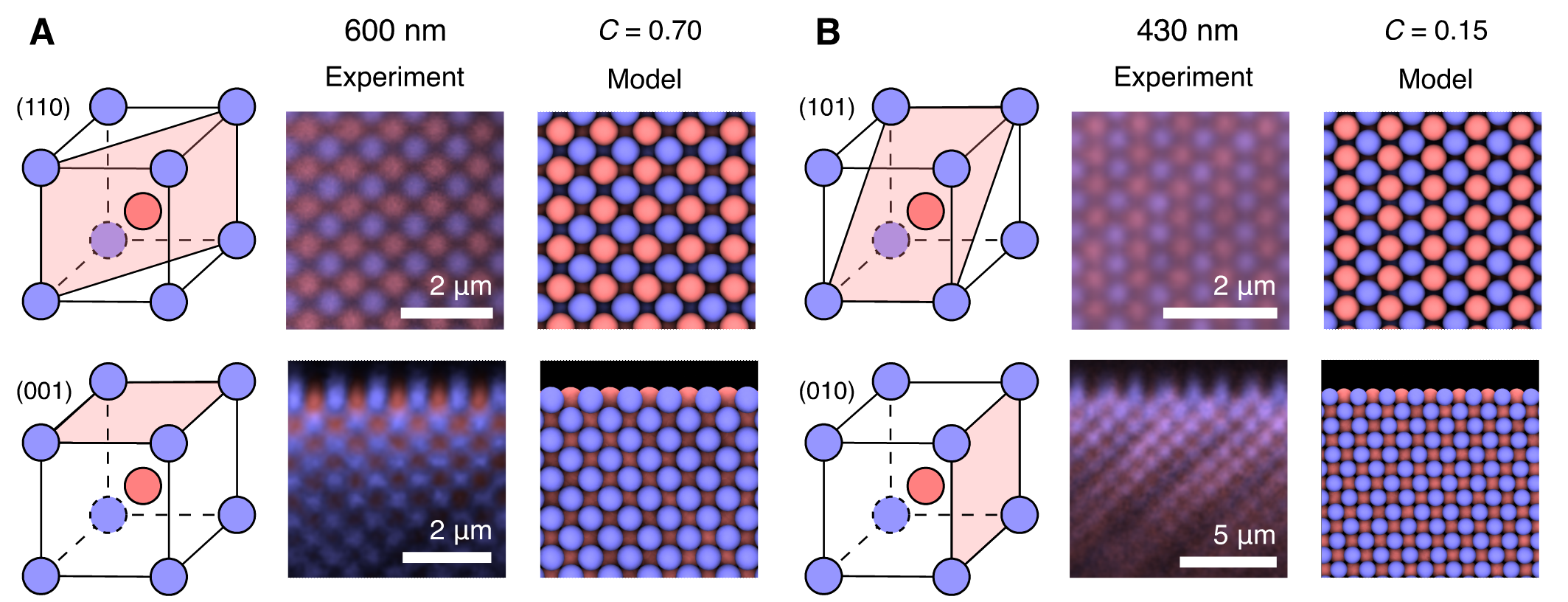}
  \caption{Confocal images, along with the corresponding best-fit model images, for the presented facets (top), as well as a vertical slice into the crystal (bottom) for: (A) a 600-nm-diameter particle crystal and (B) a 430-nm-diameter particle crystal. Each pair of images comes from the same crystal. Model images are generated from measuring and fitting the $G(r)$ to different BCT structures. In both cases, the a single BCT structure describes both the (110) and (001) crystal planes. Note that for both (A) and (B) the the experimental images for the (001) and (010) planes come from a projection of four layers of the crystal.}
  \label{SI:fig-zcharacterization}
\end{figure}

We use laser-scanning confocal microscopy to image several layers into the crystal to check the crystalline order of our structures in the third dimension. We first take an X-Y scan to image the presented facet of the crystal. We then record an X-Z scan to visualize the interior of the crystal. Because there is an index mismatch between the particles (roughly $n=1.59$) and the solvent (roughly $n=1.33$), we are only able to image ten or so layers and the image quality is reduced. To increase the contrast of the particles in the X-Z image, we compute the variance of X-Z slices over four particle diameters in the Y direction. We intentionally choose either the (010) plane or the (001) plane, which both have alternating layers of a single particle type (i.e., the first plane is all A, the second plane is all B, and so forth.) These planes also have the property that subsequent layers of the same particle type are identical, with no shift in particle positions. 
Finally, we compute the $G_{AB}(r)$ for the x-y image and the x-z image, and find the best fit BCT model for each. Importantly, we find consistent crystal structures for both planes for every crystal we tested. Exemplary images of these two orthogonal planes are shown in Fig.~\ref{SI:fig-zcharacterization}A-C for the 430-nm-diameter and 600-nm-diameter particle crystals. Due to the width of the point spread function of our confocal microscope, the 250-nm-diameter particles could not be resolved in the z-direction. These measurements into the bulk of the crystals illustrate their single crystalline nature.

\subsection{Characterizing the ensembles of crystals}

To check whether or not all of the crystals formed within a single experiment have the same crystal type, we use our quantitative pipeline to characterize dozens of single crystals for each particle size. The first step in this process is to collect images of 50--100 different crystals assembled within a single crystallization experiment. To facilitate our automated crystal detection scheme described above, we only consider crystals that sit flat on the lower coverslip surface. Once the images are collected, we pass them through the pipeline described in \secref{sec:pipeline} as a single batch. After all images are analyzed, we check that both the particle finding and $G(r)$ fitting are satisfactory, and discard any images that have too few identified particles.

\begin{figure}[h]
  \centering \includegraphics[width=\textwidth]{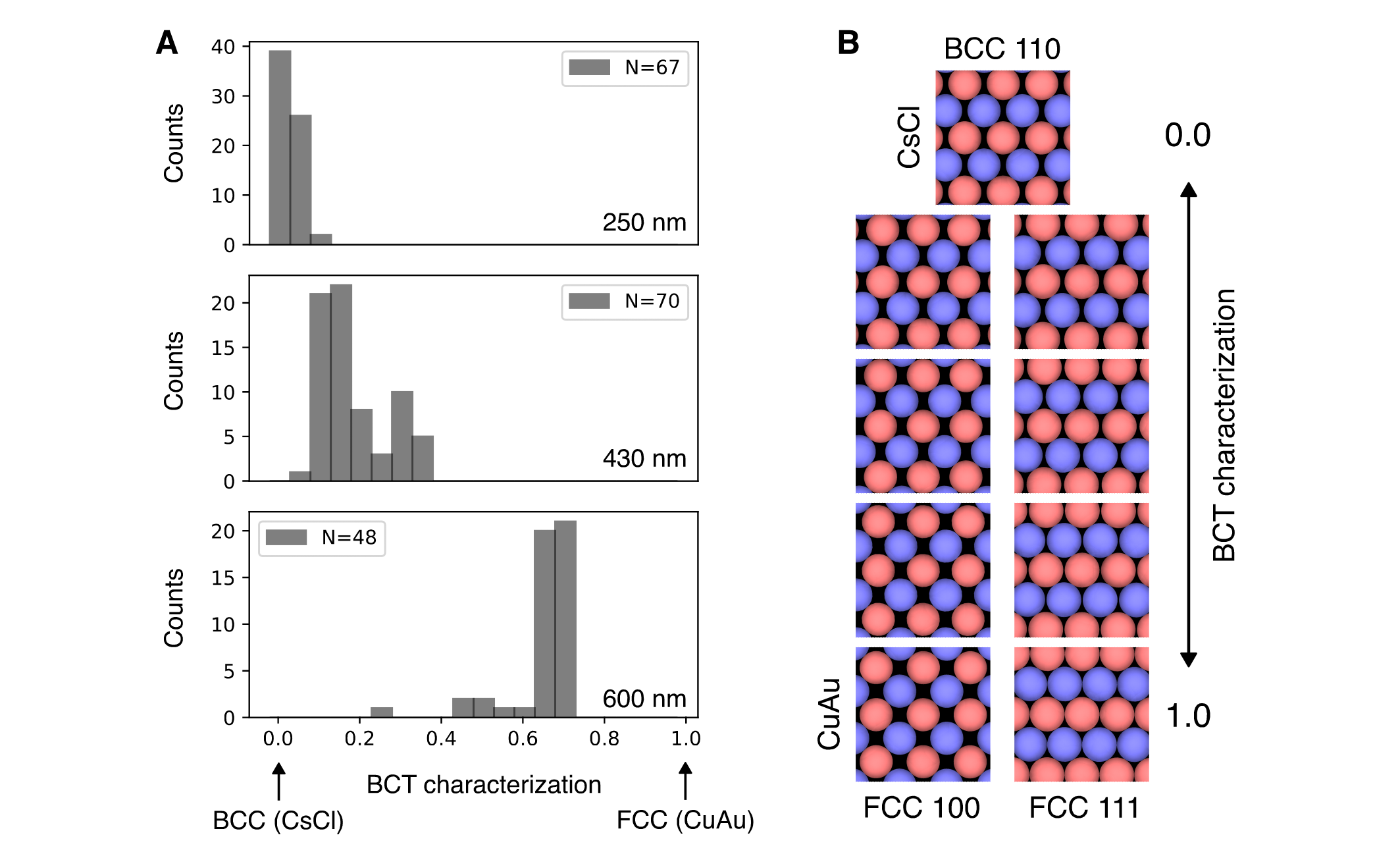}
  \caption{(A) Histograms of measured $C$-values for many crystals from each of the three different sized particles. (B) Close up of model facets for different $C$-values that characterize the BCT structure of the crystals. $C$ values of 0, 0.25, 0.5, 0.75, and 1.0 are shown in descending order.}
  \label{SI:fig-cvaluehistrogram}
\end{figure}

We find that all of the crystals within the ensemble have the same lattice type and compositional order for each particle size. Figure~\ref{SI:fig-cvaluehistrogram}A shows histograms of the best fit BCT characterization of the structures for the collected crystal images for particle diameters of 600-nm (bottom), 430-nm (middle), and 250-nm (top). For each particle type, we find consistent BCT structures for all of the crystals within the ensemble. For example, the 250-nm-diameter particles exhibit a CsCl symmetry and compositional order with mean and median values of $C=0.02\pm0.03$ and $C=0$. The 600-nm-diameter particles show a BCT structure with mean and median values of $C=0.65\pm0.09$ and $C=0.65$, that could easily be mistaken for a FCC-CuAu structure without careful quantitative analysis. The 430-nm-diameter particles yield an intermediate BCT structure with mean and median values of $C=0.18\pm0.08$ and $C=0.15$. For reference, Fig.~\ref{SI:fig-cvaluehistrogram}B shows the arrangement of particles on the most prominent facets for different BCT characterizations.

\subsection{Understanding the crystal habits}
\label{sec:habits}

Beyond the symmetry and compositional order of the crystal lattices, we also characterize and understand the crystal habits (i.e., the geometrical shapes of the single crystals). We hypothesize that the crystal habits that we observe are consistent with expectations from the Wulff construction, modified to account for the fact that crystal growth happens next to a flat surface.

\textit{Comparing crystal habits with the equilibrium Wulff construction.}
The Wulff construction is a method to determine the equilibrium shape of crystal structures that form~\cite{wulff1901xxv,auyeung2014DNA,o2016programming}. Since different crystal planes have different surface energies, the crystal habit that minimizes the total surface energy per volume in general involves a complex combination of many crystallographic planes. The Wulff construction is an algorithmic method to predict the shape that minimizes the surface energy, assuming that the bounding surface comprises only allowed crystal planes. To construct this shape, one places crystal planes around an origin point so that the minimum distance from the point to the plane is proportional to the surface energy density of that plane. The minimum enclosed volume of all of these planes is the Wulff construction. In essence, this approach creates a geometry where the exposed planes are the low-surface-energy planes.

\begin{figure}[h!]
  \centering \includegraphics[width=\textwidth]{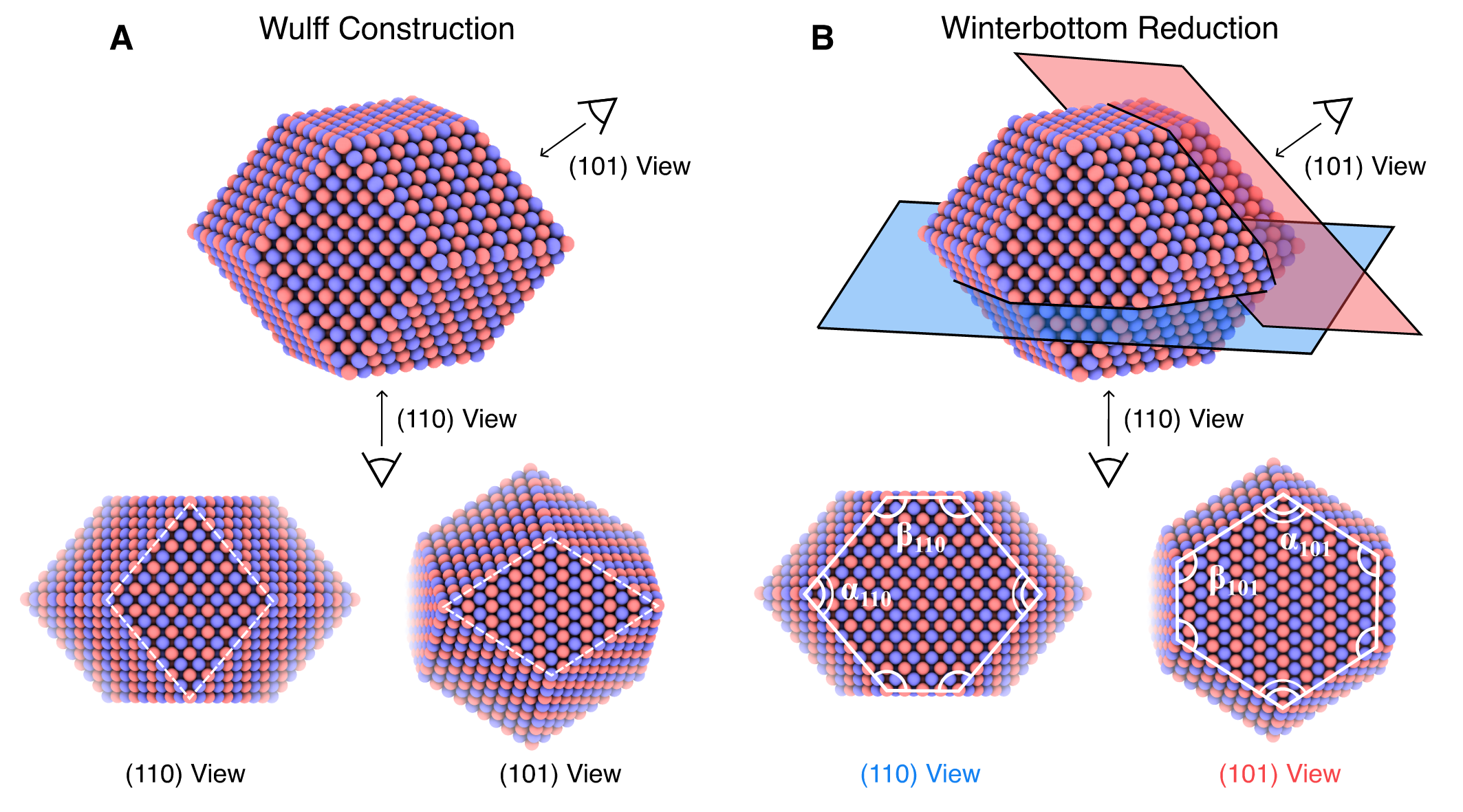}
  \caption{\textbf{Schematic for facet growth at a surface.} (A) The Wulff construction for a BCT crystal with c=0.5. Below are shown the facets from the (110) and (101) planes. These show only four sides, highlighted by the white, dashed line. (B) A cartoon of the Winterbottom-like construction. Due to the presence of a surface, growth in one direction is inhibited. We model this by removing particles from a Wulff construction below a crystal plane, shown as blue for the (110) plane and red for the (101) plane. Below are the facets shown for these two planes. Due to the removal of particles, these facets now exhibit six sides with two distinct angles, $\alpha$ and $\beta$. The precise location where the plane cuts the Wulff constrution affects the edge lengths of the exposed facet, but the angles remain fixed.}
  \label{SI:fig-winterbottomreduction}
\end{figure}

Using the Wulff construction, we create model crystals based on estimates of the relative surface-energy densities of the binary BCT crystal structure. We estimate the surface energy density by considering the unit cell of a crystal plane and computing the ratio of the number of missing bonds to the area of the unit cell; the surface energy densities for different planes are rescaled by the lowest-surface-energy density plane. Such a Wulff construction is shown in Fig.~\ref{SI:fig-winterbottomreduction}A for a BCT crystal with $C=0.5$. For $C$-values between 0 and 1, the ($\pm1\pm10$), ($\pm10\pm1$), and ($0\pm1\pm1$) planes have the lowest surface energies and have Wulff shapes that are distorted rhombic dodecahedra, which are stretched along the z-axis. 

\textit{Accounting for interactions with the coverslip during late-stage growth.}
However, when we look at the (110) or (101) facets of these shapes, we find that they always have four sides, whereas the crystal facets that we find in experiment always have six sides. We hypothesize that this difference comes from the proximity of the coverslip and its effect on late-stage growth.
To test this hypothesis, we carry out an analysis based on the Winterbottom construction~\cite{winterbottom1967equilibrium}, which we discuss next.

In the main text, we describe how we grow crystals either in droplets or in a bulk sample. In both of these cases, as the crystals increase in mass, they settle either at the bottom of the droplet, or against the bottom coverslip of the sample chamber. When this occurs, the crystal will preferentially grow from its exposed sides. This situation is reminiscent of the Winterbottom construction, where a crystal nucleates and grows preferentially from a surface~\cite{winterbottom1967equilibrium, lewis2020single}. In our case, the substrate-facet surface energy is not relevant because the particles do not actually adhere to the interface.

To emulate the Winterbottom construction, we can take a crystal generated by the Wulff construction and pass a plane through it that we imagine being parallel to the substrate. Figure~\ref{SI:fig-winterbottomreduction}B  shows an example with planes slicing through the (110) and (101) facets. This additional plane mimics the surface that the crystal is resting upon, so we reduce the crystal by removing all particles beneath the plane. If we perform this Winterbottom reduction for the (110) or (101) facets and look at the reduced face, we see they now have six sides. On each of these facets there are two distinct angles, $\alpha$ and $\beta$, which can therefore be used as another way to quantify the BCT structure of the crystal, since these angles change monotonically with $C$.

\begin{figure}[h!]
  \centering \includegraphics[width=\textwidth]{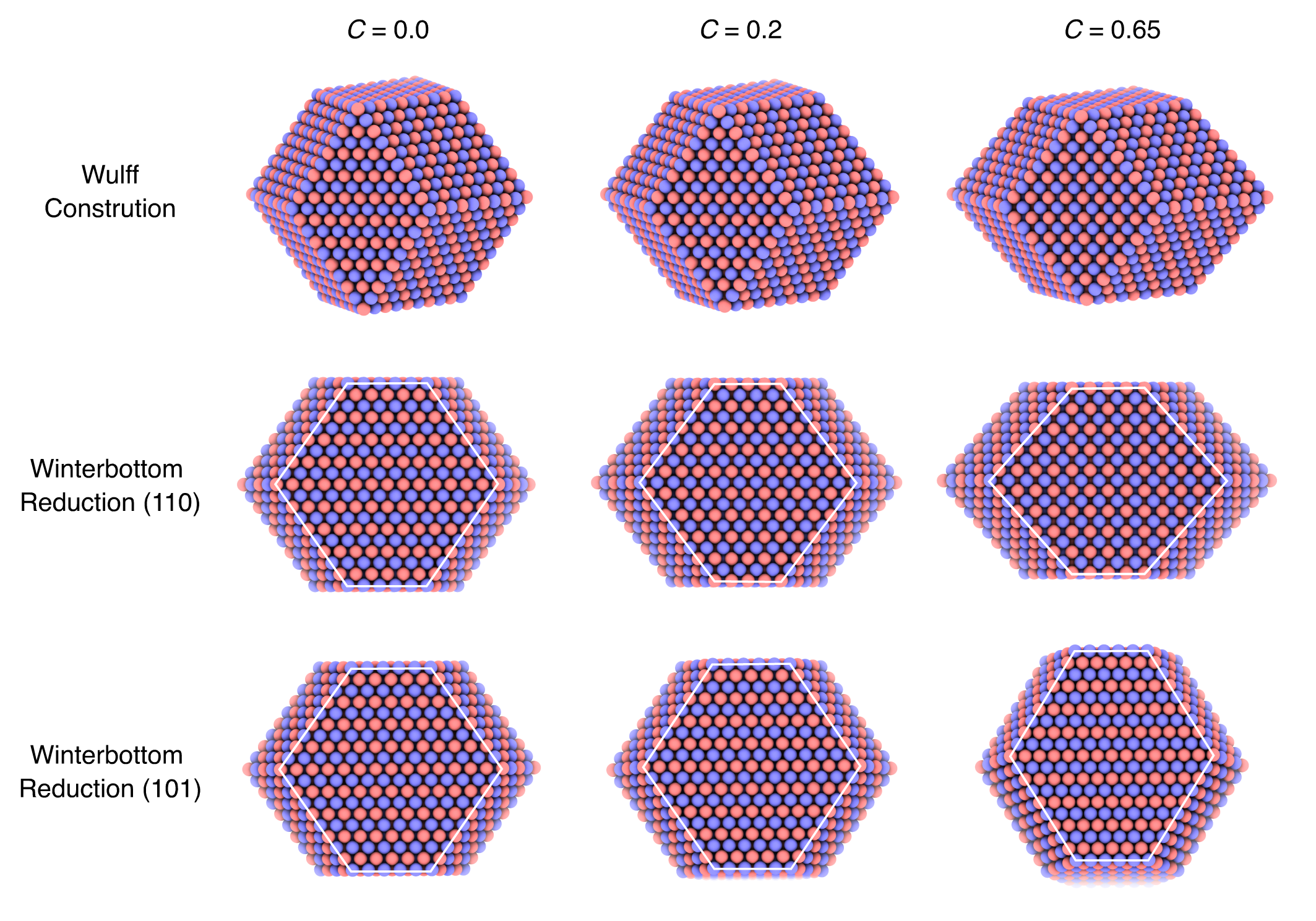}
  \caption{This shows the expected Wulff constructions and Winterbottom-like constructions of the (110) and (101) planes for the average $C$-values from the 250-nm ($C$=0), 430-nm ($C$=0.2) and 600-nm ($C$=0.65) crystals.}
  \label{SI:fig-averageWinterbottomFacets}
\end{figure}

Remarkably, despite the simplicity of the our approach, we find that the Winterbottom reduction captures the shape of our observed crystal facets, as well as their internal angles. In Figs.~\ref{SI:fig-600nmcrystal}A-~\ref{SI:fig-250nmcrystal}A, the model crystal facets generated from the $G(r)$ fitting and Winterbottom Reduction show the same facet angles as the experimental system. For the 600-nm, 430-nm, and 250-nm-diameter particles, we find that each particle size have an average $c$-value that describes the ensemble of measured structures. The corresponding Wulff constructions and Winterbottom reduction are shown in Fig.~\ref{SI:fig-averageWinterbottomFacets}, and the predicted facet angles for the (110) or (101) planes are shown in Table~\ref{SI:table-facetAngles}. Importantly, the crystals synthesized by seeded growth also showed the same facet angles. Thus, we conclude that interactions with the coverslip can account for the observed differences between the expected equilibrium Wulff shapes and the observed (110) and (101) facets in our experiments.

\begin{table}[h!]
    \centering
    \begin{tabular}{c| c c | c c}
    d (nm) & $\alpha_{110}$ & $\beta_{110}$ & $\alpha_{101}$ & $\beta_{101}$ \\
    \hline
    250 & 109.5$\pm$0.4 & 125.3$\pm$0.4 & 109.5$\pm$0.4 & 125.3$\pm$0.4\\
    430 & 106.2$\pm$1.6 & 126.9$\pm$0.8 & 111.1$\pm$0.8 & 124.4$\pm$0.4 \\
    600 &  97.0$\pm$1.9 & 131.5$\pm$1.0 & 116.0$\pm$1.1 & 122.0$\pm$0.5 
    \end{tabular}
    \caption{Based on the average measured BCT structure for each particle size, this table lists the expected range of angles for (110) and (101) facets seen in experiment.}
    \label{SI:table-facetAngles}
\end{table}

\subsection{Rationalizing the observed structural colors}

Figure 4D,E in the main text shows examples of colloidal crystals assembled from 250-nm-diameter and 430-nm-diameter particles that show red and green coloration, respectively, under normal incidence. We rationalize the colors that we observe by considering Bragg diffraction from the crystalline lattices~\cite{asher2004diffraction}. Therefore, the wavelength in vacuum of the diffracted light should be approximately
\begin{equation}
  \lambda = \frac{2 n_c d_{(110)}}{n},
\end{equation}
where $n_c$ is the effective refractive index of the crystal, $n$ is the diffraction order, and $d_{(110)}$ is the distance between (110) planes of CsCl. For a perfect CsCl crystal, $d_{(110)}=2\sqrt{2/3}r$ for a particle radius $r$.

We hypothesize that the 430-nm-diameter particle crystal shows green coloration, whereas the 250-nm-diameter particle crystal shows red, due to second-order diffraction. Assuming the red coloration of the 250-nm-diameter particle crystal is due to the primary ($n=1$) Bragg reflection, we can estimate the effective refractive index of the colloidal crystal. Taking $\lambda\approx 650$~nm, we find a value of the effective refractive index of $n_c=1.59$, which is approximately equal to the refractive index of polystyrene. This number seems reasonable given that roughly 70\% of the crystal volume is composed of polystyrene. Substituting $n_c=1.59$ and a particle radius $r=430/2$~nm into our expression for the wavelength of diffracted light yields roughly $\lambda=1100$~nm. While light of this color is squarely in the infrared, the second order diffracted light would have a wavelength of $\lambda=550$~nm, which is in the green portion of the visible spectrum, as we observe in our experiments. Therefore, we conclude that the coloration of the 430-nm-diameter particle crystals is likely due to second order diffraction, but this explanation would need to confirmed by direct experimental measurement to know definitively.

\section{Theoretical analysis of self-assembly in microfluidic droplets}

In this section, we describe the theoretical model used to predict the probability of assembling a single-domain crystal within a droplet.
We introduce the formal definition of $\taug$ and show how it can be calculated from a dynamical model of droplet-confined self-assembly based on classical theories of nucleation and growth.
We then specialize the discussion to the case of discrete-step temperature-ramp experiments and show that the temperature step used in the experiments presented in the main text, $\Delta T = 0.1^\circ$C, is near optimal in this system.

\subsection{Predicting the probability of single-crystal self-assembly in droplets}

We first consider an arbitrary monotonic annealing protocol, $T(t)$.
The probability that a single crystal assembles from an initial nucleation event occurring at temperature $T$ is given by
\begin{equation}
  p_{\text{1x}}\big(T;T(t)\big) = p_{\text{first-nuc}}\big(T;T(t)\big) \times p_{\text{no-sec-nuc}}\big(T;T(t)\big).
\end{equation}
The first term, $p_{\text{first-nuc}}$, is the probability that the initial nucleation event occurs at the temperature $T$ and is given by
\begin{equation}
  \label{eq:first-nuc}
  p_{\text{first-nuc}}\big(T;T(t)\big) = \left|\frac{dT}{dt}\right|^{-1}\! \Vdrop k_{\text{n}}\big(\rho_0,T\big) \exp\left[-\int_T^{T(0)} dT'\, \left|\frac{dT}{dt}\right|^{-1}\! \Vdrop k_{\text{n}}\big(\rho_0,T'\big)\right],
\end{equation}
where $\Vdrop$ is the droplet volume, $\kn$ is the nucleation rate density, and $\rho_0$ is the initial uniform colloid concentration.
In the case of an isothermal protocol, this reduces to $p_{\text{first-nuc}}(T) = 1 - \exp[-\Vdrop \kn(\rho_0,T)t_{\text{max}}]$, where $t_{\text{max}}$ is the duration of the experiment.
If the annealing protocol begins above the melting temperature and ends at a temperature where there is no nucleation barrier, then the first nucleation event must happen at some time during the protocol; this means that $p_{\text{first-nuc}}[T(t)] \equiv \int_{T(t_{\text{max}})}^{T(0)} dT\, p_{\text{first-nuc}}(T;T(t)) = 1$.
The second term, $p_{\text{no-sec-nuc}}$, is the probability that no subsequent nucleation events occur in the droplet.
This probability depends on the growth dynamics of the crystal and its effect on the colloid concentration field, $\rho(r,t)$, elsewhere in the droplet.
Assuming that the initial nucleation event occurs near the center of the droplet, we can approximate this term as
\begin{equation}
  p_{\text{no-sec-nuc}}\big(T;T(t)\big) = \exp\left[-\int_{T(t_{\text{max}})}^T dT'\,\left|\frac{dT}{dt}\right|^{-1} \int_{\Vdrop} d\bm{r}\, \kn\big(\rho(r,t),T'\big)\right].
\end{equation}
Finally, we can integrate over $T(t)$ to calculate $p_{\text{1x}}$ for the complete protocol:
\begin{equation}
  p_{\text{1x}}[T(t)] \equiv \int_{T(t_{\text{max}})}^{T(0)} dT\,p_{\text{1x}}\big(T;T(t)\big) = \int_{T(t_{\text{max}})}^{T(0)} dT\,p_{\text{first-nuc}}\big(T;T(t)\big) \times p_{\text{no-sec-nuc}}\big(T;T(t)\big).
\end{equation}

Both $p_{\text{first-nuc}}$ and $p_{\text{no-sec-nuc}}$ are incredibly sensitive to the temperature due to their dependence on the nucleation rate density, $\kn$.
It is therefore useful to introduce a characteristic growth time, $\taug$, which represents the time required for all subsequent nucleation events to be suppressed.
In the general case of a monotonic annealing protocol, we define $\taug$ according to
\begin{equation}
  \label{eq:taug}
  \int_0^{\taug} dt\, \Vdrop \kn\big(\rho_0,T(t_{\text{first-nuc}}+t)\big) = \int_{T(t_{\text{max}})}^{T(t_{\text{first-nuc}})} dT'\,\left|\frac{dT}{dt}\right|^{-1} \int_{\Vdrop} d\bm{r}\, \kn\big(\rho(r,t),T'\big),
\end{equation}
where $t_{\text{first-nuc}}$ is the time at which the first nucleation event occurred.
It turns out that $\taug$ varies slowly with $T_{\text{first-nuc}} \equiv T(t_{\text{first-nuc}})$, which allows for further approximations as discussed below.
It is illuminating to consider the example of an isothermal protocol, in which case $\taug$ is given by
\begin{equation}
  \label{eq:taug-isothermal}
  \taug = \int_0^\infty dt\, \Vdrop^{-1} \int_{\Vdrop} d\bm{r}\,\frac{\kn\big(\rho(r,t),T\big)}{\kn\big(\rho_0,T\big)} = \int_0^\infty dt\, \left\langle \frac{\kn\big(\rho(r,t),T\big)}{\kn(\rho_0,T)}\right\rangle_{\Vdrop}.
\end{equation}
(The upper limit on the time integral can be extended to infinity if $\taug \ll t_{\text{max}}$.)
In this case, the ratio of nucleation rates, averaged over the droplet, decays to zero once equilibrium is established.
The single-crystal probability can then be written as $p_{\text{1x}}(T) = \{1 - \exp[-\Vdrop \kn(\rho_0,T)t_{\text{max}}]\} \exp[-\Vdrop\kn(\rho_0,T)\taug]$.
In the more general case of a time-dependent protocol, $\taug$ has an analogous physical meaning, although the definition given above cannot be simplified in the same way.
In what follows, we discuss how $\rho(r,t)$ can be calculated from the dynamical growth model, and we provide an illustrative example calculation for an isothermal protocol.
We then demonstrate that $\taug$ is only weakly dependent on $T_{\text{first-nuc}}$, as claimed above.

\subsection{Calculating $\tau_{\text{g}}$ from the dynamical growth model}

We can calculate $\taug$ from the dynamical growth model presented in Reference~\cite{Hensleye2114050118}.
This model accounts for both reaction-limited nucleation kinetics, which are dependent on the rate at which DNA-grafted colloids roll or slide over one another, and diffusion-limited growth.
Using the equilibrium and kinetic measurements performed in Reference~\cite{Hensleye2114050118}, we describe the crystal--vapor equilibrium and nucleation rate density by the equations
\begin{align}
  \log [\rho_{\text{eq}}(T) / \text{M}] &= (10.50 / ^\circ\text{C}) T - 573.39 \\
  \kn(\rho,T) &= (e^{22.4}\,\text{M}^{-1}\text{s}^{-1} / 10^{-13}\,\text{m}^3) \rho \exp[16\pi\gamma(T)^3 / 3(\log S)^2] \\
  \gamma(T) &= -(0.972 / ^\circ\text{C}) T + 51.85,
\end{align}
where $\rho_{\text{eq}}(T)$ is the equilibrium colloidal gas density at temperature $T$, $S \equiv \rho / \rho_{\text{eq}}$ is the supersaturation, and $\gamma(T)$ is the surface tension of the crystal--vapor interface, multiplied by the colloid diameter squared.
We then use the Wilson--Frenkel growth law, the continuity boundary condition at the crystal--vapor interface, and the conservation of colloids in the droplet to obtain the deterministic growth equations for the crystal radius, $R(t)$,
\begin{align}
  \frac{dR}{dt} &= \frac{\alpha\alpha_{\text{diff}}}{\alpha + \alpha_{\text{diff}}} \left(\frac{v_\infty}{S}\right) \left[\sigma_\infty - 2 \left(\frac{\xi}{R}\right)\right] \label{eq:dRdt} \\
  \sigma_\infty &= \left[1 - \frac{3\alpha}{2(\alpha + \alpha_{\text{diff}})} \left(\frac{R}{R_{\text{drop}}}\right)\right]^{-1} \left[\frac{\rho_0 - \rho_{\text{eq}}}{\rho_{\text{eq}}} - \frac{\rho_{\text{c}}}{\rho_{\text{eq}}}\left(\frac{R}{R_{\text{drop}}}\right)^3 - \frac{3\alpha}{(\alpha + \alpha_{\text{diff}})}\left(\frac{\xi}{R_{\text{drop}}}\right)\right] \\
  \alpha &= \frac{\kappa / v_\infty}{1 + K(S=0)^{-1} + \kappa / v_\infty} \\
  v_\infty &= D \rho^{2/3}, \quad \alpha_{\text{diff}} = \frac{\rho_{\text{eq}} D S}{\rho_{\text{c}} v_\infty R}, \quad \rho_{\text{c}} = \frac{6\phi}{\pi d^3}, \quad \xi = \left(\frac{R^*}{R}\right)^2 \gamma, \quad R^* = \frac{2\gamma\rho_{\text{eq}}}{\rho_0 - \rho_{\text{eq}}},
\end{align}
with constants
\begin{equation}
  D = 10^{-12}\,\text{m}^2\text{s}^{-1}, \quad
  d = 6\times10^{-7}\,\text{m}, \quad
  \phi = 0.74, \quad
  \kappa = 1\,\text{s}^{-1}, \quad
  \text{and} \quad
  K(S=0) = 0.01.
\end{equation}
Here, $\rho_0$ is the initial colloid number density, $D$ is the self-diffusion coefficient, $\phi$ is the packing fraction in the crystal phase, and $\kappa$ is the typical time required for a colloid to diffuse its own diameter while rolling or sliding on another colloid.
The equilibrium constant $K$ describes the attachment of a single colloid to a growing nucleus, which gives rise to an attachment coefficient, $\alpha$, that is less than unity at low supersaturations, as described in Reference~\cite{Hensleye2114050118}.
Note that we have modified the Gibbs--Thomson term involving $\xi$ such that it decays rapidly once the crystal grows larger than $R^*$; this modification is necessary for the vapor phase in the finite-size droplet to reach equilibrium and can be justified by noting that the crystal rapidly develops planar interfaces beyond the critical nucleus size.
Also note that if the annealing protocol is time-dependent, then the quantities $\rho_{\text{eq}}$ and $\gamma$ are time-dependent as well.

\begin{figure}[h]
  \centering\includegraphics[width=\textwidth]{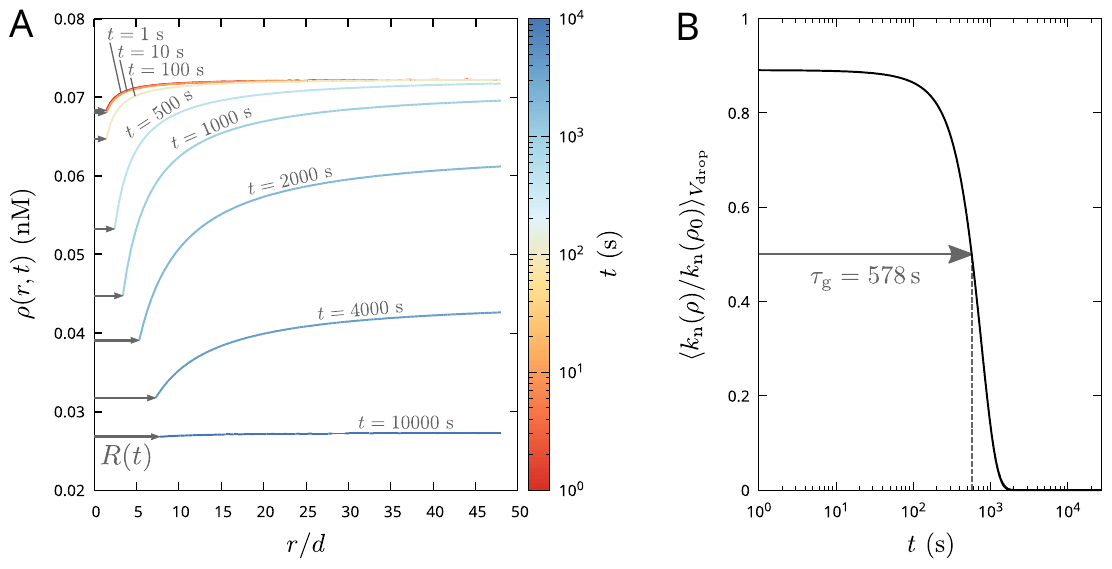}\vskip-1ex
  \caption{\label{fig:example-isothermal}
    \textbf{(A)} An example concentration profile measured from the center of the droplet as a function of the time post nucleation [i.e., $R(t=0) = 1.1 R^*$] in an isothermal protocol [$T = 52.3\,^\circ\text{C}$, $\Vdrop = 10^{-13}\,\text{m}^3$, and $\rho_0 = 0.5\%\,(\text{v/v})$].  The inner boundary condition for the concentration profile is located at the crystal--vapor interface, $R(t)$.  \textbf{(B)} The corresponding calculation of $\taug$.  Note that the ratio of nucleation rates starts below unity since, according to the deterministic growth model, the critical nucleus itself is sufficient to establish a non-uniform concentration field.  Comparison of panels A and B shows that the average nucleation rate across the droplet volume does not decrease substantially until the concentration profile spreads out across the entire droplet and the outer boundary condition begins to decrease from its initial value, $\rho_0$.}
\end{figure}

\begin{figure}[h]
  \includegraphics[width=\textwidth]{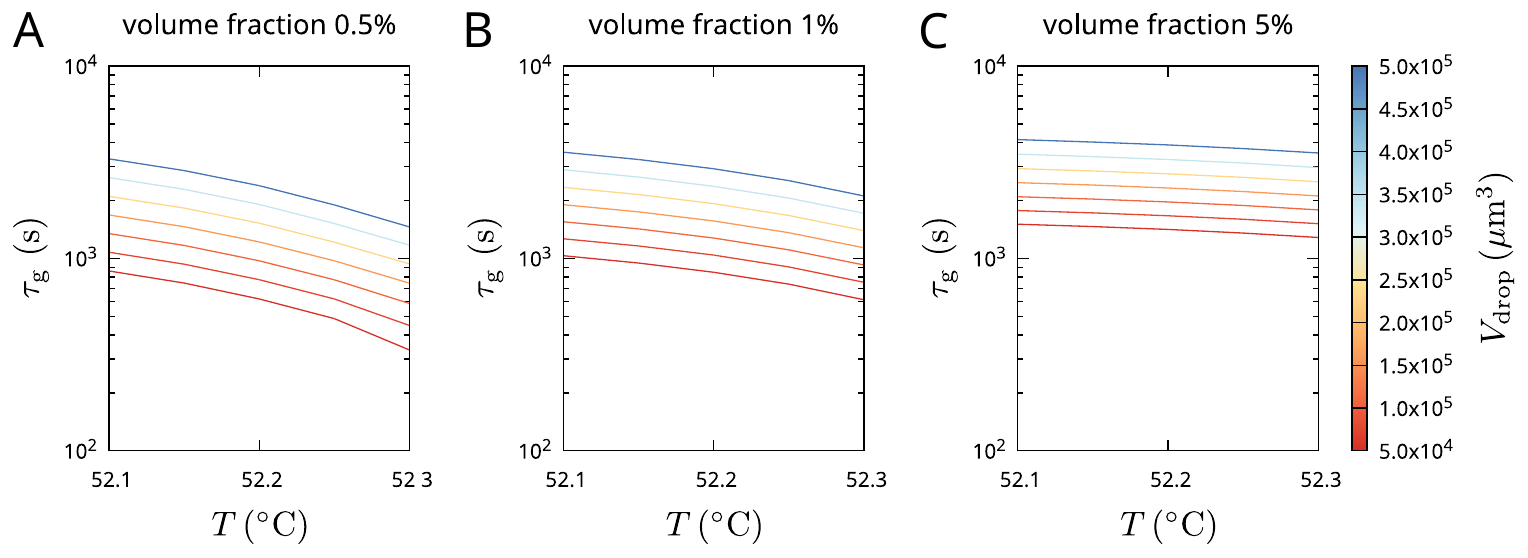}\vskip-1ex
  \caption{\label{fig:taug-isothermal}
    Compared to $\kn$, which varies by orders of magnitude over a $0.1^\circ\text{C}$ temperature range, calculations using the deterministic growth model show that $\taug$ is relatively insensitive to temperature.  Panels show example calculations assuming an initial particle volume fraction of \textbf{(A)} 0.5\%, \textbf{(B)} 1\%, and \textbf{(C)} 5\%.}
\end{figure}

We solve the differential equation for the crystal radius numerically using the Euler method starting from an initial radius 10\% larger than the critical radius, $R^*$.
We then compute the time-dependent concentration field, $\rho(r,t)$, from the numerical solution to the differential equation for $R(t)$,
\begin{equation}
  \sigma(r,t) \equiv \frac{\rho(r,t) - \rho_{\text{eq}}}{\rho_{\text{eq}}} = \sigma_\infty - \frac{\alpha}{\alpha + \alpha_{\text{diff}}}\left[\sigma_\infty - 2 \left(\frac{\xi}{R}\right)\right] \left(\frac{R}{r}\right), \quad R < r < R_{\text{drop}}.
  \label{eq:concentration-field}
\end{equation}
With expressions for $\rho(r,t)$ and $\kn(\rho,T)$, we can evaluate $\taug$ using the definition given in \eqref{eq:taug}.
Example calculations of the time-dependent concentration field and $\taug$ for an isothermal protocol are shown in \figref{fig:example-isothermal}.
Importantly, this example shows that the average nucleation rate across the entire droplet does not decrease substantially from its initial value until the concentration profile spreads out to reach the edge of the droplet at $r=R_{\text{drop}}$.
Thus, at times shorter than $\taug$, regions within the droplet volume that are far from the initial nucleation site do not feel any effects of the growing crystal.
We also show that $\taug$ is indeed weakly dependent on the temperature, despite $\kn$ varying by orders of magnitude over a temperature range of $\sim 0.1\,^\circ\text{C}$, in \figref{fig:taug-isothermal}.

\subsection{Predicting the probability of single-crystal self-assembly in discrete-step temperature-ramp droplet experiments}
\label{sec:discrete-step}

We now apply the theory developed above to predict the probability of assembling a single crystal in a discrete-step temperature-ramp experiment.
Since we do not know the exact temperature at which the protocol is initiated [i.e., $T(0)$] in any particular experimental run, we approximate the expected value of $p_{\text{1x}}[T(t)]$ by assuming a continuous linear ramp with constant ramp rate $dT/dt$.
The probability that the first nucleation event occurs at the temperature $T$ in this approximation is thus
\begin{equation}
  \label{eq:first-nuc-continuous}
  p_{\text{first-nuc}}\big(T;T(t)\big) = \left|\frac{dT}{dt}\right|^{-1}\! \Vdrop\kn(\rho_0,T) \exp\left[-\left|\frac{dT}{dt}\right|^{-1}\! \int_T^\infty dT'\, \Vdrop\kn(\rho_0,T')\right].
\end{equation}
However, it is crucial that the computation of $p_{\text{no-sec-nuc}}$ captures the stepwise nature of the protocol, since each step of the protocol is essentially isothermal over a time period, $\Delta t$, that is much longer than $\taug$.
We therefore compute the probability of a second nucleation event using the isothermal expression under the assumption that $\taug \ll \Delta t$,
\begin{equation}
  \label{eq:T-first-nuc}
  p_{\text{no-sec-nuc}}\big(T;T(t)\big) = \exp[-\Vdrop \kn(\rho_0,T) \taug(T)].
\end{equation}
We can now take advantage of the weak temperature dependence of $\taug$ by solving the growth model only at the most probable initial nucleation temperature.
We do so by computing the expected value of the initial nucleation temperature,
\begin{equation}
  \langle T_{\text{first-nuc}} \rangle_{T(t)} = \int_0^\infty dT\,T p_{\text{first-nuc}}(T;T(t)),
\end{equation}
and then evaluating $\taug^*$ at $\langle T_{\text{first-nuc}} \rangle_{T(t)}$ using the isothermal expression, \eqref{eq:taug-isothermal}.
The final expression for the single crystal probability is
\begin{equation}
  \label{eq:p1x-step}
  p_{\text{1x}}[T(t)] = \left|\frac{dT}{dt}\right|^{-1}\! \int_0^\infty dT\, \Vdrop\kn(\rho_0,T) \exp\left[-\left|\frac{dT}{dt}\right|^{-1}\! \int_T^\infty dT'\, \Vdrop\kn(\rho_0,T') - \Vdrop \kn(\rho_0,T) \taug^* \right].
\end{equation}

Fig.~2 in the main text shows the comparison between the predictions made using this approach and the results of discrete-step temperature-ramp experiments.
Importantly, the predictions do not involve any adjustable parameters.
The overestimation of $p_{\text{1x}}$ at the slowest ramp rates is likely due to partial evaporation of the droplets, which tends to increase the supersaturation more quickly than the rate specified by the temperature protocol alone.
The fact that this effect is greatest in the smallest droplets is consistent with this hypothesis, since the first nucleation event is expected to occur at lower temperatures (and thus later times in a protocol) in these cases.

\subsection{Comparing the probability of secondary nucleation events in discrete-step and continuous temperature-ramp droplet experiments}

\begin{figure}[h]
  \centering \includegraphics[width=0.9\textwidth]{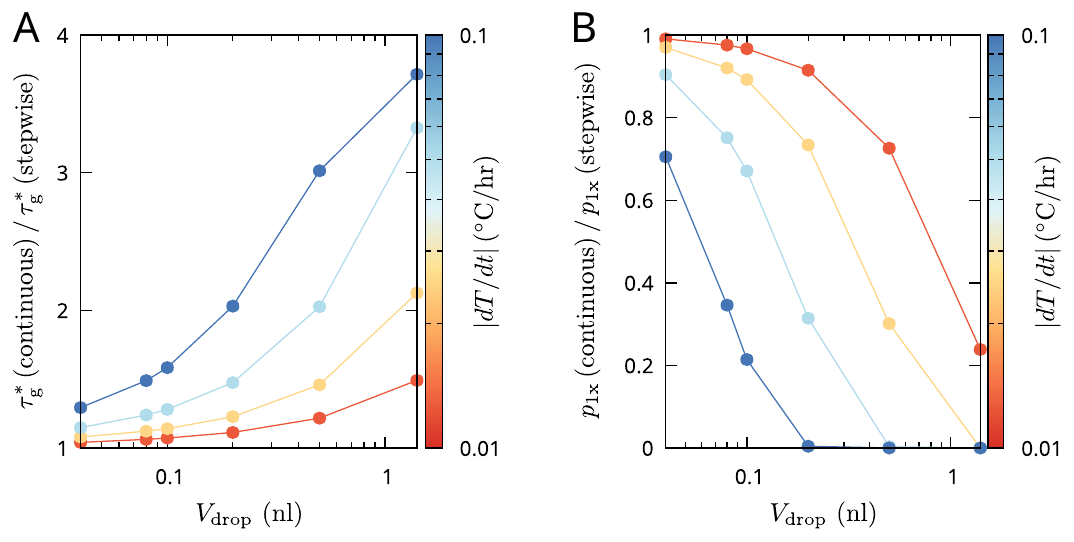}\vskip-1ex
  \caption{\label{fig:continuous}
    Comparison of \textbf{(A)} $\tau_{\text{g}}^*$ and \textbf{(B)} the predicted single crystal fraction resulting from continuous versus stepwise temperature ramps with the same average ramp rate, $dT/dt$, assuming an initial colloid volume fraction of 0.5\%.}
\end{figure}

This theory predicts that the stepwise nature of the temperature ramps used in our experiments substantially increases the probability of assembling a single crystal.
For the sake of comparison, we can consider what happens when the temperature ramp is continuous.
In this case, we must use \eqref{eq:taug} to compute $\taug^*$, assuming that the initial nucleation event occurs at $\langle T_{\text{first-nuc}} \rangle_{T(t)}$ and that the temperature continues to decrease linearly in time after $t_{\text{first-nuc}}$.
Note that it is still reasonable to use $\taug^*$ instead of computing $\taug(T)$, since $\taug$ remains weakly dependent on the initial nucleation temperature in the case of a continuous temperature ramp.
The probability of assembling a single crystal using such a protocol is thus
\begin{equation}
  \label{eq:p1x-continuous}
  p_{\text{1x}}[T(t)] = \left|\frac{dT}{dt}\right|^{-1}\! \int_0^\infty dT\, \Vdrop\kn(\rho_0,T) \exp\left[-\left|\frac{dT}{dt}\right|^{-1}\! \int_{T-|dT/dt|\taug^*}^\infty dT'\, \Vdrop\kn(\rho_0,T') \right].
\end{equation}
\figref{fig:continuous} shows that $p_{\text{1x}}$ is always predicted to be lower using a continuous protocol than it is using a stepwise protocol with the same average ramp rate, assuming that $\taug \ll \Delta t$.
Furthermore, the relative performance of the continuous protocol decreases as the droplet volume and the ramp rate increase.
These effects can be understood by considering the ratio of the $\taug^*$ values for the continuous and stepwise protocols, which is always greater than unity.
In the case where the temperature continues to decrease continuously following the initial nucleation event, the homogeneous nucleation rate far from the initial nucleus actually \textit{increases} before the nonuniform concentration profile is able to spread out across the entire droplet via diffusion.
The net result is that it takes longer in a continuous protocol for the growth of the first crystal to suppress nucleation elsewhere in the droplet, particularly when the droplet is large and/or the ramp rate is fast.

\subsection{Predicting an optimal temperature-ramp protocol}

The analysis above shows that discrete steps of duration $\taug \ll \Delta t$ tend to increase $p_{\text{no-sec-nuc}}$ relative to a continuous protocol.
However, this theory cannot hold for all protocols with discrete steps of arbitrary duration.
On the one hand, a continuous protocol can be represented in practice by a discrete protocol with the step height, $\Delta T$, and width, $\Delta t$, made extremely small.
On the other hand, a protocol with arbitrarily large $\Delta T$ is essentially a deep quench, which we know does not lead to single-crystal assembly.
This raises the question of what the optimal step height should be when designing a discrete-step temperature-ramp protocol at a fixed average ramp rate.

To address this question, we revisit the key assumptions made in developing the theory for the discrete-step protocol in \secref{sec:discrete-step}.
It is important to note that the absolute temperature at which the ramp starts, $T(0)$, is a relevant parameter; in other words, we need to specify $\Delta T$, $|dT/dt| = |\Delta T / \Delta t|$, and $T(0)$ to define a discrete-step temperature ramp precisely.
However, because we typically do not know $T(0)$ to sufficient accuracy in an experiment, it is only practical to predict a range of possible outcomes for protocols whose initial temperature is randomly selected from a uniform distribution.
This is the strategy taken in the calculations below.

\begin{figure}[h]
  \centering \includegraphics[width=\textwidth]{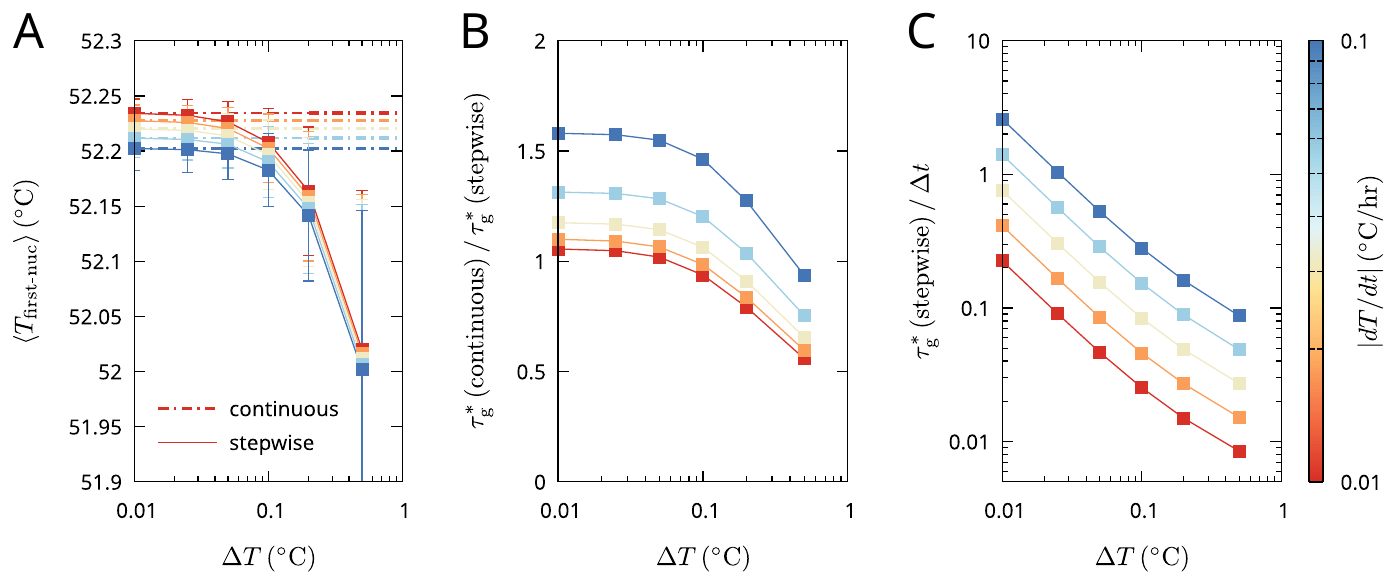}\vskip-1ex
  \caption{\label{fig:discrete}
    Comparisons of \textbf{(A)} first-nucleation temperatures and \textbf{(B)} $\taug^*$ for continuous and stepwise protocols with varying step heights, $\Delta T$.  The error bars indicate the standard deviation assuming a uniform distribution of starting temperatures.  \textbf{(C)} Comparison of $\taug^*$ to the step width, $\Delta t$, for varying step heights and ramp rates.}
\end{figure}

First, we consider the accuracy of the continuous approximation for $p_{\text{first-nuc}}$, \eqref{eq:first-nuc-continuous}, used in the comparison between discrete-step and continuous ramps above.
\figref{fig:discrete}A shows the predicted temperatures at which the first nucleation event occurs in discrete and continuous protocols as a function of $\Delta T$ for selected ramp rates.
In the case of the predictions for discrete protocols, the points and error bars indicate the mean and standard deviation, respectively, of $\langle T_{\text{first-nuc}} \rangle$ for a uniform distribution of starting temperatures, $T(0) \in [T_0, T_0 + \Delta T)$, where $T_0$ is a temperature above the melting temperature.
The initial nucleation temperatures tend to be lower when using discrete protocols, although the variations resulting from differences in $T(0)$ can be quite large.
This comparison indicates that the continuous approximation, \eqref{eq:first-nuc-continuous}, is likely to be sufficiently accurate [i.e., within the uncertainty due to $T(0)$] for step heights up to approximately $0.1\,^\circ\text{C}$.
In this range of $\Delta T \lesssim 0.1\,^\circ\text{C}$, we find that $\taug^*$, which is computed at the average expected first-nucleation temperature $\langle T_{\text{first-nuc}} \rangle$, is indeed greater for discrete protocols than for continuous protocols, as shown in \figref{fig:discrete}B.
However, the values of $\taug^*$ for the discrete protocols tend to decrease at large step heights due to the decrease in the mean $\langle T_{\text{first-nuc}} \rangle$.

Second, we consider the suitability of the isothermal approximation for $p_{\text{no-sec-nuc}}$ used in \eqref{eq:p1x-step}.
This approximation is only valid when $\taug^* \ll \Delta t$.
As shown in \figref{fig:discrete}C, this condition is met when $\Delta T$ is large and the ramp rate is slow.
This approximation is thus likely to be of sufficient accuracy for step heights larger than about $0.05\,^\circ\text{C}$.

Taken together, the example calculations reported in \figref{fig:discrete} indicate that \eqref{eq:p1x-step}, which treats the initial nucleation probability using a continuous temperature ramp and the secondary nucleation probability using an isothermal protocol, is applicable only if the temperature step height is on the order of $0.05$ to $0.1\,^\circ\text{C}$.
If the step height is smaller than this range (i.e., the protocol is closer to being continuous), then crystal growth cannot be considered to be isothermal, and probability of secondary nucleation increases.
Yet if the step height is larger than this range (i.e., the protocol is less continuous), then the first nucleation event is more likely to occur at a low temperature, which implies a faster initial nucleation rate and thus also promotes secondary nucleation.
Thus, in addition to \eqref{eq:p1x-step} being most applicable in this range of step heights, we also expect these stepwise protocols to yield single crystals with the highest probability given an imposed ramp rate (or, equivalently, a prescribed maximum duration of the experiment).
Nonetheless, we should keep in mind that there is more inherent variability in the results of discrete-step temperature-ramp protocols due to the uncertainty in the absolute starting temperature, $T(0)$, which means that the actual yield may vary slightly from experiment to experiment.

\section{Theoretical analysis of seeded growth experiments}

\subsection{Identifying isolated seeds and diffusion-limited growth}

Seeded growth experiments are carried out using monodisperse, pre-assembled seeds in a bath of `weak' colloidal particles at constant supersaturation.
The annealing conditions are chosen such that the weak particles are only moderately supersaturated, leading to growth of the crystalline seeds without homogeneous nucleation of new crystals from the weak-particle bath.
Analogously to our discussion of late-stage crystal growth within droplets, we expect the growth of isolated seeds in bulk solution under these conditions to be diffusion-limited.
Starting from an initial spherical seed containing $N_0$ particles, we integrate \eqref{eq:dRdt} in the limit $\alpha \gg \alpha_{\text{diff}}$ and $R \gg \xi$ to obtain the deterministic diffusion-limited growth law
\begin{equation}
  \label{eq:growth}
  N(t) = \left\{ \frac{2}{3} \left[ \frac{3}{2} N_0^{2/3} + \frac{2\pi d D}{\phi^{1/3}} \left(\rho_0 - \rho_{\text{eq}}\right) t \right]\right\}^{3/2},
\end{equation}
where $\rho_0$ and $\rho_{\text{eq}}$ are the initial and equilibrium concentrations, respectively, of the weak particles in the bulk phase, and time is measured from the beginning of the seeded growth experiment.
At long times, when $N(t) \gg N_0$, this expression predicts the scaling law $N(t) \sim t^{3/2}$.

\begin{figure}[t]
  \centering \vskip-2ex \includegraphics[width=\textwidth]{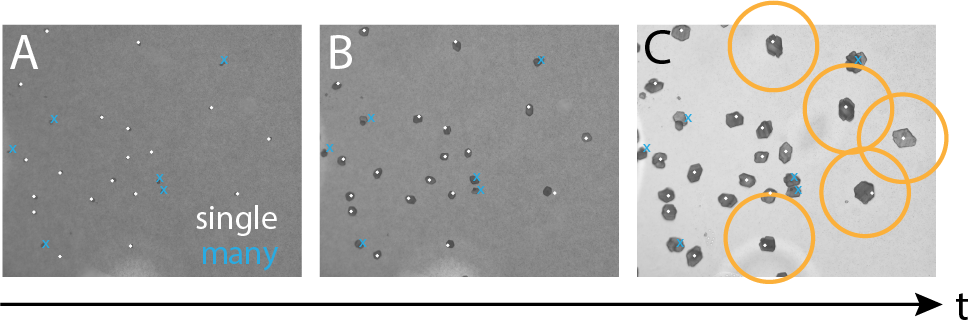}\vskip-1ex
  \caption{\label{fig:isolated}
  In seeded growth experiments, a high local concentration of seeds can reduce the growth rate of nearby seeds. (A) We first distinguish between single-crystal seeds (white dots) and polycrystal seeds (blue crosses). (B--C) Crystals grown from single-crystal seeds are then classified as being isolated if the initial seed is separated from all other seeds by a distance of at least 3 times the diameter of the crystal at the conclusion of the seeded growth experiment, as indicated by the yellow circles on the right. The isolated crystals are also the largest crystals in the final frame, indicating that they grew the fastest as described in the text.}
\end{figure}

Fig.~3C in the main text shows that the scaling law predicted for diffusion-limited growth at constant supersaturation is borne out by the data.
However, we find that the prefactor dictating the specific rate of the growth can decrease if the seeds are too close together.
To understand how far apart seeds must be placed in order to grow at the maximum rate, we consider the concentration profile predicted by \eqref{eq:concentration-field}.
In the limit $\alpha \gg \alpha_{\text{diff}}$ and $R \gg \xi$ relevant to growth in the weak-particle bath, the decrease in the local particle concentration due to the growth of a single isolated crystal decays inversely with the ratio $r / R(t)$, where $r$ is the distance from the center of the crystal and $R(t)$ is the current radius of the crystal.
We therefore predict that a crystal will be effectively isolated for the entirety of the seeded growth experiment as long as it is separated from all other seeds by a distance of at least $3R(t_{\text{max}})$, where $t_{\text{max}}$ is the duration of the experiment.
This choice ensures that the far-field reduced chemical potential difference, $\Delta\mu / k_{\text{B}}T \simeq S - 1$, of crystals identified as being isolated never drops below $2/3$ of the bulk value.
As demonstrated in Fig.~3C of the main text, this criterion for identifying effectively isolated seeds allows us to select crystals that maintain a monodisperse size distribution during seeded growth.
This criterion is also illustrated in \figref{fig:isolated}, which shows a mixture of isolated and non-isolated crystals within the same field of view.

\subsection{Comparison with isothermal nucleation and growth in bulk solution}
\begin{figure}[h!]
  \centering \includegraphics[width=0.6\textwidth]{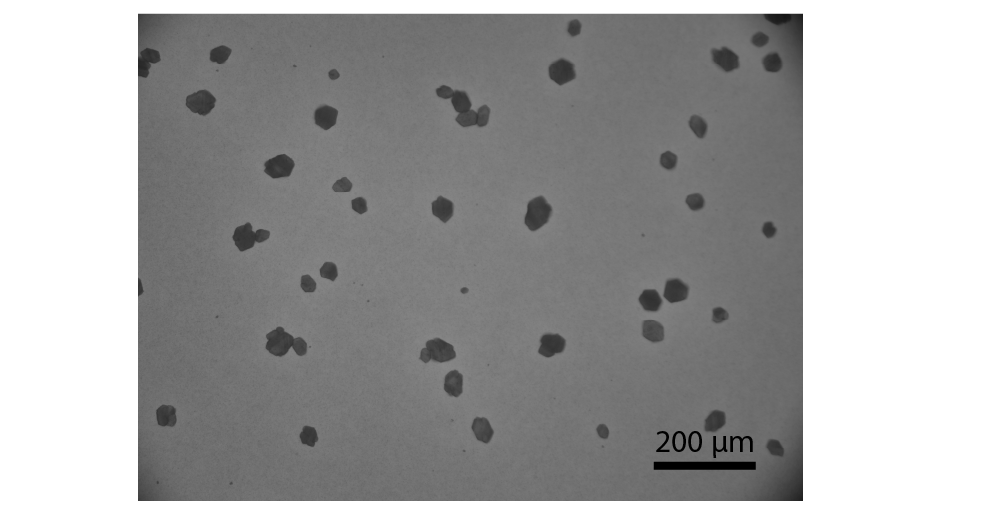}\vskip-1ex
  \caption{\label{fig:bulk}
  An image of crystals formed by isothermal nucleation and growth in bulk that was compared to the data from the seeded growth experiment in Fig.~3C of the main text.}
\end{figure}

To emphasize the surprisingly narrow size distribution of isolated crystals grown using our two-step protocol, we compare our final crystal sizes with the distribution of crystal sizes obtained from an isothermal experiment in the inset of the main text Fig.~3C.
This bulk-nucleation experiment was analyzed at a time when some crystals were roughly the size as the isolated crystals from the seeded experiment; the image used in this analysis is shown in \figref{fig:bulk}.
The differences between the bulk and seeded nucleation experiments can be understood by noting that uniform supersaturation of a bulk fluid leads to stochastic nucleation.
Because the nucleation times are exponentially distributed, the time periods over which the various crystals can grow vary widely.
The resulting distribution of final crystal sizes is therefore extremely broad.
The formation of polycrystals is also apparent in \figref{fig:bulk}, since the initial nucleation events also occur randomly in space.

~\\~\\~\\~\\~

\section{Supplementary Figures}

\begin{figure}[!h]
  \centering \includegraphics{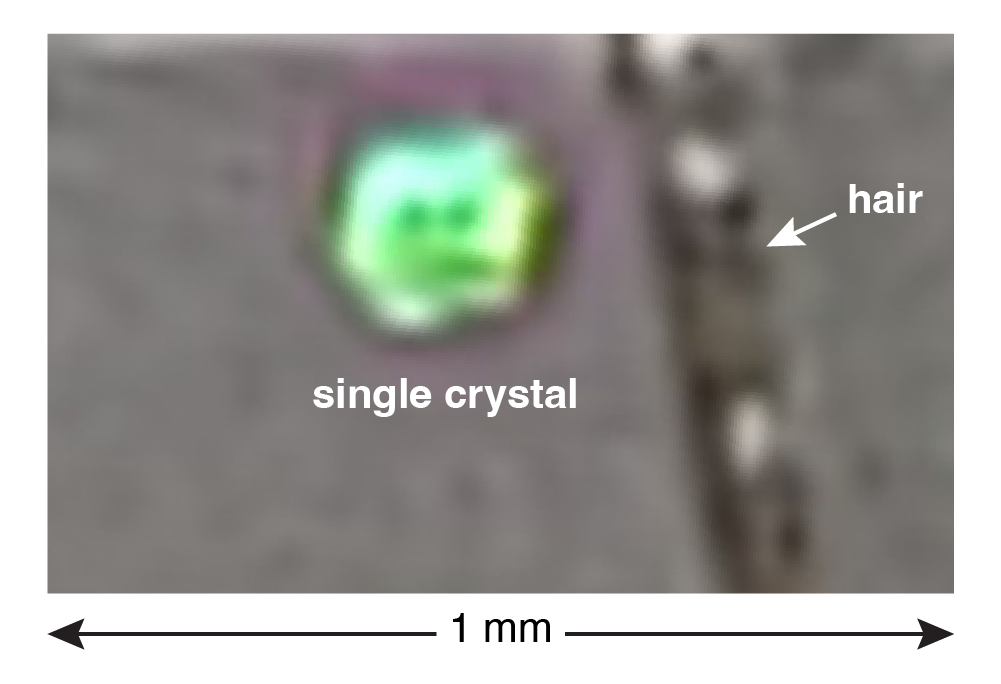}\vskip-1ex
  \caption{\label{fig:SI-Cellphone}
  A cellphone image of a capillary containing a large single crystal formed from 400-nm-diameter particles next to a human hair taken without any intermediate optics. The width of the image is 1~mm.
  }
\end{figure}

\bibliography{SI}